\documentclass[12pt]{article}
\usepackage{geometry} % see geometry.pdf on how to lay out the page. 
\usepackage{amssymb,amsmath} % or letter or a5paper or ... etc
\usepackage{cite}
\def\Bbb{\mathbb}
\def\BZ{\Bbb Z} \def\BR{\Bbb R}
\def\BC{\Bbb C}  
\def\BN{\Bbb N}\def\BH{\Bbb H}

\def\Fg{\mathfrak{g}}
\def\Fh{\mathfrak{h}}
\catcode`@=11 \@addtoreset{equation}{section} \catcode`@=12

\begin{document}
\begin{titlepage}
\renewcommand{\thefootnote}{\fnsymbol{footnote}}
\noindent
{\tt IITM/PH/TH/2008/2}\hfill
{\tt arXiv:0807.4451v3}\\
{\tt IMSc-2008-07-13} \hfill March 2009 \textbf{[version 3.11]} \\
%\vspace{1.0cm}
\begin{center}
\large{\bf  
Generalized Kac-Moody Algebras from CHL dyons}
\end{center} 
\bigskip 
\begin{center}
Suresh Govindarajan\footnote{\texttt{suresh@physics.iitm.ac.in}} \\
\textit{Department of Physics, Indian Institute of Technology Madras,\\ Chennai 600036, INDIA.}
\\[5pt]
and \\[5pt]
K. Gopala Krishna\footnote{\texttt{gkrishna@imsc.res.in}} \\
\textit{The Institute of Mathematical Sciences,\\
CIT Campus, Taramani, 
Chennai 600113 INDIA}
\end{center}

\begin{abstract}
We provide evidence for the existence of a family of generalized 
Kac-Moody(GKM) superalgebras, $\mathcal{G}_N$, whose 
Weyl-Kac-Borcherds denominator formula gives rise to a genus-two 
modular form at level $N$, $\Delta_{k/2}(\mathbf{Z})$, for 
$(N,k)=(1,10)$, $(2,6),$ $(3,4)$, and possibly $(5,2)$. The 
square of the automorphic form is the modular transform of the 
generating function of the degeneracy of CHL dyons in asymmetric 
$\BZ_N$-orbifolds of the heterotic string compactified on $T^6$.  
The new generalized Kac-Moody superalgebras all arise as 
different `automorphic corrections' of the same Lie algebra and 
are closely related to a generalized Kac-Moody superalgebra 
constructed by Gritsenko and Nikulin. The automorphic forms, 
$\Delta_{k/2}(\mathbf{Z})$, arise as additive lifts of Jacobi 
forms of (integral) weight $k/2$ and index $1/2$.  We note that 
the orbifolding acts on the imaginary simple roots of the 
unorbifolded GKM superalgebra, $\mathcal{G}_1$ leaving the real 
simple roots untouched. We anticipate that these superalgebras 
will play a role in understanding the `algebra of BPS states' in 
CHL compactifications.
\end{abstract}
\end{titlepage}
\section{Introduction}

The advent of D-branes and in particular, the successful 
microscopic description of the entropy of supersymmetric black 
holes by Strominger and Vafa\cite{Strominger:1996sh} has provided 
enormous impetus to the counting of BPS states in a variety of 
settings. Among these, the $\mathcal{N}=4$ supersymmetric 
theories, have provided a fairly robust playground for testing 
and developing new ideas. The high degree of supersymmetry 
ensures the existence of non-renormalization theorems as well as 
the existence of dualities. This paper focuses on 
four-dimensional superstring compactifications, the CHL 
orbifolds, with $\mathcal{N}=4$ supersymmetry. Such theories 
typically have a triality of descriptions as heterotic and type 
IIA/IIB string theories\cite{Duff:1995sm}.

The spectrum of $1/2$-BPS states is independent of moduli -- 
there is no `jumping' phenomenon. In a remarkable 
paper\cite{Dijkgraaf:1996it}, Dijkgraaf, Verlinde and Verlinde 
proposed that the degeneracy of $1/4$-BPS states in the heterotic 
string compactified on $T^6$ is generated by a Siegel modular 
form of weight $10$. More precisely, the 
degeneracy $d(n,\ell,m)$ is given by\footnote{We choose a normalization
for the modular forms that is natural to the Weyl denominator formula.
For instance, our $\Delta_5(\mathbf{Z})$ is $1/64$ of the same form
defined by Gritsenko and Nikulin\cite{Nikulin:1995}. Further, 
$64=2^{\frac{12}{2}}$ also happens to be the size of a $1/4$ BPS multiplet 
of states.}
\[
\frac{64}{\Phi_{10}(\mathbf{Z})} = \sum_{(n,\ell,m)>0} d(n,\ell,m)\ q^n
r^\ell s^m
\eqno{\eqref{DVVformula}}
\]
where $\mathbf{Z}\in \BH_2$, the Siegel upper-half space and 
$(n,\ell,m)=(\frac12\mathbf{q_e}^2, \mathbf{q_e}\cdot 
\mathbf{q_m}, \frac12\mathbf{q_m}^2)$ are T-duality invariant 
combinations of electric and magnetic charges. A key feature of 
the formula is that it is also S-duality invariant i.e., the 
modular form is invariant under an $SL(2,\BZ)$ suitably embedded  in
$Sp(2,\BZ)$. In ref. \cite{Nikulin:1995}, Gritsenko and Nikulin 
have shown the existence of a generalized Kac-Moody (GKM) 
superalgebra, $\mathcal{G}_1$, whose Weyl-Kac-Borcherds 
denominator formula gives rise to a modular form (with character) 
of weight 5, $\Delta_5(\mathbf{Z})$, which squares to give 
$\Phi_{10}(\mathbf{Z})$. It has been anticipated by Harvey and 
Moore\cite{Harvey:1995fq,Harvey:1996gc} that the algebra of BPS 
states must form an algebra and one suspects that the GKM 
superalgebra, $\mathcal{G}_1$, must play such a role in this 
example. However, this correspondence has not been fully realised 
for this example(see ref. \cite{Cheng:2008fc} however for some recent 
progress). One of our motivations has been to understand the 
relation of the algebra of BPS states to this GKM superalgebra. 

A family of asymmetric orbifolds of the heterotic string 
compactified on $T^6$ leads to heterotic compactifications that 
preserve ${\mathcal N}=4$ supersymmetry but the gauge symmetry is 
of reduced rank\cite{Chaudhuri:1995fk}. These are called the CHL 
compactifications. In ref. \cite{Jatkar:2005bh}, Jatkar and Sen, 
constructed a family of genus-two modular forms, 
$\Phi_k(\mathbf{Z})$, at level $N$ that play a role analogous to 
$\Phi_{10}$, for $\BZ_N$-orbifolds with $N=2,3,5,7$ and 
$(k+2)=24/(N+1)$. The modular group is a subgroup of $Sp(2,\BZ)$ 
reflecting the fact that the orbifolding breaks the S-duality 
group to the sub-group, $\Gamma_1(N)$, of 
$SL(2,\BZ)$\cite{Sen:2005pu}. The main result of this paper is to 
provide evidence the existence of a family of GKM superalgebras, 
$\mathcal{G}_N$, by showing the existence of modular forms, 
$\Delta_{k/2}(\mathbf{Z})$, which:  (i) square to 
$\Phi_k(\mathbf{Z})$ and (ii) appear as the denominator formula 
for $\mathcal{G}_N$ for $N=2,3,5$. These constructions parallel 
the construction of a family of GKM algebras as $\BZ_N$-orbifolds 
of the fake Monster Lie algebra\cite{Borcherds:1990} by 
Niemann\cite{Niemann}.
 
We will show that the modular forms $\Delta_{k/2}(\mathbf{Z})$ 
are indeed given by the denominator formula for GKM superalgebras, 
$\mathcal{G}_N$, that 
are closely related to the GKM superalgebra constructed by 
Gritsenko and Nikulin (we call the superalgebra $\mathcal{G}_1$) 
from the modular form $\Delta_5(\mathbf{Z})$. In particular, we 
observe
\begin{enumerate}
\item All the algebras arise as (different) automorphic 
corrections to the Lie algebra $\Fg(A_{1,II})$ associated with a 
rank three Cartan matrix $A_{1,II}$. In other words, one has 
$\Fg(A_{1,II})\subset \mathcal{G}_N$.
\item The real simple roots (and hence the Cartan matrix $A_{1,II}$) for the 
$\mathcal{G}_N$ are identical to the real roots of 
$\Fg(A_{1,II})$. This implies that the Weyl group is identical as 
well\footnote{However, for $N>1$, this Weyl group is no longer the symmetry group of the lattice of dyonic charges as it was for $N=1$. The reason is that the lattice of dyonic charges is \textit{not} generated by $1/\Phi_k(\mathbf{Z})$. Of course, as was shown by Jatkar  and Sen\cite{Jatkar:2005bh}, it is another closely related modular form $1/\widetilde{\Phi}_k(\mathbf{Z})$ that generates the lattice of dyonic charges and their degeneracies.}.
\item The multiplicities of the imaginary simple roots are, 
however, different. For instance, imaginary roots of the form $t 
\eta_0$, where $\eta_0$ is a primitive light-like simple root, 
have a multiplicity $m(t \eta_0)$ given by the formula:
\begin{displaymath}
1-\sum_{t\in \BN} m(t \eta_0)\ q^n 
= \prod_{n\in\BN} (1-q^n)^{\tfrac{k-4}2} (1-q^{Nn})^{\tfrac{k+2}2}
\end{displaymath}
Note that this formula correctly reproduces the answer given for 
$\mathcal{G}_1$ by Gritsenko and Nikulin\cite{Nikulin:1995}.
\item There are also other imaginary simple roots which are not 
light-like whose multiplicities are determined implicitly by the 
modular form $\Delta_{k/2}(\mathbf{Z})$.
\end{enumerate}

\noindent The organization of the paper is as follows: After the 
introductory section, section 2 provides the necessary physical 
background on CHL strings and the counting of $1/4$-BPS dyons in 
the theory. Section 3 provides the necessary mathematical 
background leading to the Weyl-Kac-Borcherds denominator formula 
for GKM algebras. In section 4, which contains the main results 
of the paper, we provide evidence for the existence of Siegel 
modular forms with character, $\Delta_{k/2}(\mathbf{Z})$, at 
level $N$ as the additive lift of weak Jacobi forms of 
half-integral index. In sections 4.2, 4.3 and 4.4, we show that 
these modular forms arise as the Weyl-Kac-Borcherds denominator 
formula for GKM superalgebras, $\mathcal{G}_N$, with imaginary 
simple roots whose multiplicities given by the 
$\Delta_{k/2}(\mathbf{Z})$. In section 4.5, we attempt to provide 
a physical interpretation for the weak Jacobi forms of 
half-integral index. We conclude with a summary of our results 
and a discussion in section 5. Several mathematical appendices 
have been included. In particular, we review the use of weak 
Jacobi forms in the construction of Siegel modular forms in 
appendix B. In appendix C, we discuss the additive lift of Jacobi 
forms with index $1/2$ at level $N$. This generalizes an existing 
result at level $1$ and proves the modularity of the forms 
$\Delta_{k/2}(\mathbf{Z})$ that we have constructed. Appendix D 
gives the explicit Fourier expansion for the modular forms 
$\Delta_{k/2}(\mathbf{Z})$ that were used in arriving at the 
results presented in section 4.

\section{Background on CHL strings}

The heterotic string compactified on $T^6$ and its asymmetric 
$\BZ_N$ orbifolds provide us with four-dimensional 
compactifications with $\mathcal{N}=4$ supersymmetry. Writing 
$T^6$ as $T^4 \times \widetilde{S}^1\times S^1$, consider the 
$\BZ_N$ orbifold given by the transformation corresponding to a 
$1/N$ unit of shift in $\widetilde{S}^1$ and a simultaneous 
$\BZ_N$ involution of the Narain lattice of signature $(4,20)$ 
associated with the heterotic string compactified on $T^4$. This 
leads to the CHL compactifications of interest in this 
paper\cite{Jatkar:2005bh}. The heterotic string compactified on $T^4 
\times \widetilde{S}^1\times S^1$ is dual to the type IIA string 
compactified on $K3 \times \widetilde{S}^1\times S^1$. The 
$(4,20)$ lattice gets mapped to $H^*(K3,\BZ)$ in the type IIA 
theory and the orbifolding $\BZ_N$ is a Nikulin involution combined
with 
the $1/N$ shift of $\widetilde{S}^1$.

\noindent The low-energy theory consists of the following bosonic fields: 
\begin{itemize}
\item[(i)] the $\mathcal{N}=4$ supergravity multiplet with the 
graviton, a complex scalar, $S_H$ and six graviphotons; and
\item[(ii)] $m=\left([48/(N+1)]-2\right)$  $\mathcal{N}=4$ vector multiplets 
each containing a gauge field and six scalars.
\end{itemize}
In terms of the variables that appear in the heterotic description, the bosonic
part of the low-energy effective action (up to two derivatives)
is\cite{Schwarz:1993vs,Cvetic:1995uj,Lerche:1999ju}
\begin{multline}\label{lowenergy}
S=\int d^4x \sqrt{-g}\left[R - \frac{\partial_\mu S_H\ 
\partial^\mu \bar{S}_H}{2~ \textrm{Im}(S_H)^2}  
+\frac18 \textrm{Tr}(\partial_\mu ML\ \partial^\mu M L) \right. \\
\left.-\frac14 \textrm{Im}(S_H)\ F_{\mu\nu}LML\ F^{\mu\nu}
+\frac14 \textrm{Re}(S_H)\ F_{\mu\nu}L\ \widetilde{F}^{\mu\nu} 
\right]\ ,
\end{multline} 
where $L$ is a Lorentzian metric with signature $(6,m)$, $M$ is a 
$(6+m)\times (6+m)$ matrix valued scalar field satisfying $M^T=M$ 
and $M^TLM=L$ and $F_{\mu\nu}$ is a $(6+m)$ dimensional vector field 
strength of the $(6+m)$ gauge fields.

The moduli space of the scalars is given by 
\begin{equation}
\big(\Gamma_1(N)\times SO(6,m;\BZ)\big)\Big\backslash\! 
\left(\frac{SL(2)}{U(1)} \times \frac{SO(6,m)}{SO(6)\times SO(m)}\right)
\end{equation}
$SO(6,m;\BZ)$ is the T-duality symmetry and $\Gamma_1(N)\subset SL(2,\BZ)$ is
the S-duality symmetry that is manifest in the equations of motion and is
compatible with the charge quantization\cite{Sen:2005pu}. The fields that 
appear at low-energy
can be organized into multiplets of these various symmetries. 
\begin{enumerate}
\item The heterotic dilaton combines with the axion (obtained by 
dualizing the antisymmetric tensor) to form the complex scalar 
$S_H$. Under $S$-duality, $S_H \rightarrow (a S_H + b)/(c S_H + 
d)$.
\item The $(6+m)$ vector fields transform as a 
$SO(6,m;\mathbb{Z})$ vector. Thus, the electric charges $\mathbf{q}_e$ 
(resp. magnetic charges $\mathbf{q}_m$) associated with these 
vector fields are also vectors (resp. co-vectors) 
of $SO(6,m,\mathbb{Z})$. Further, 
the electric and magnetic charges\footnote{To be precise, $\mathbf{q}_e$
and $-L\mathbf{q}_m$ form the doublet. Also note that the Lorentzian
inner product is
$\mathbf{q}_e^2 = \mathbf{q}_e^T L \mathbf{q}_e$. We shall not indicate
the appearance of $L$ in subsequent formulae that appear in the paper.}
transform as a doublet under 
the $S$-duality group, $\Gamma_1(N)$, where $\Gamma_1(N)$ is the 
sub-group of $SL(2,\BZ)$ matrices $\left(\begin{smallmatrix}a & b 
\\ c & d \end{smallmatrix}\right)$ with $a=d=1\mod N$ and 
$c=0\mod N$).
\end{enumerate}

One can form three T-duality invariant scalars, $\mathbf{q}_e^2$, 
$\mathbf{q}_m^2$ and $\mathbf{q}_e\cdot\mathbf{q}_m$ from the 
charge vectors. These transform as a triplet of the S-duality 
group. Equivalently, we can write the triplet as a symmetric 
matrix:
\begin{equation}
\mathcal{Q}\equiv  \left(\begin{array}{cc}\mathbf{q}_e^2 
& -\mathbf{q}_e\cdot\mathbf{q}_m \\ 
-\mathbf{q}_e\cdot\mathbf{q}_m& \mathbf{q}_m^2\end{array}\right)
\end{equation}
The $S$-duality transformation now is $\mathcal{Q}\rightarrow A 
\cdot \mathcal{Q} \cdot A^T$ with $A=\left(\begin{smallmatrix}a & 
b \\ c & d \end{smallmatrix}\right)\in \Gamma_1(N)$. The charges 
are quantized such that $N \mathbf{q}_e^2,\ \mathbf{q}_m^2 \in 2 
\BZ$ and $\mathbf{q}_e\cdot\mathbf{q}_m\in \BZ$. There exist many 
more invariants due to the discrete nature of  the T-duality 
group\cite{Banerjee:2007sr} for $N=1$ and more appear when $N>1$.

\subsection{BPS multiplets}

Four-dimensional compactifications with $\mathcal{N}=4$ 
supersymmetry admit two kinds of BPS states: (i) $1/2$-BPS 
multiplets that preserve eight supercharges (with $16$ states in 
a multiplet) and (ii) $1/4$-BPS multiplets that preserve four 
supercharges(with $64$ states in a multiplet). 
The masses of the $1/4$-BPS states are 
determined in terms of their charges by means of the BPS 
formula\cite{Cvetic:1995uj,Duff:1995sm,Lerche:1999ju}:
\begin{multline}
\label{1/4-BPS formula}
\left(M^2_\pm\right)_{1/4-BPS} = \frac1{S_H -\bar{S}_H} 
\Big[(\mathbf{q}_e+S_H\mathbf{q}_m)^T (M+L) 
(\mathbf{q}_e+\bar{S}_H \mathbf{q}_m) \\
\pm\ \frac12\ \sqrt{(\mathbf{q}_e^T (M+L)\mathbf{q}_e)(\mathbf{q}_m^T
(M+L)\mathbf{q}_m) -(\mathbf{q}_e^T (M+L)\mathbf{q}_m)^2}\ \Big]\ .
\end{multline}
The square of the mass of a $1/4$-BPS state is  $\textrm{max}(M^2_+,M^2_-)$. 
$1/2$-BPS states appear when  the electric and magnetic 
charges are parallel (or anti-parallel) i.e., 
$\mathbf{q}_e\propto L \mathbf{q}_m$. The BPS mass formula for $1/2$-BPS states
can be obtained as a specialization of the $1/4$-BPS mass formula given above. 
When $\mathbf{q}_e\propto L \mathbf{q}_m$, the terms inside the 
square root appearing in the $1/4$-BPS mass formula vanish leading to 
the $1/2$-BPS formula
\begin{equation}
\label{halfBPS formula}
\left(M^2\right)_{1/2-BPS} = \frac1{S_H -\bar{S}_H} 
\Big[(\mathbf{q}_e+S_H\mathbf{q}_m)^T (M+L) 
(\mathbf{q}_e+\bar{S}_H \mathbf{q}_m) \Big]\ .
\end{equation}

\subsection{Counting $1/2$-BPS states}

We will now consider the counting of purely electrically charged $1/2$-BPS
states.
Such electrically charged states are in one to one 
correspondence with the states of the CHL orbifold of the 
heterotic string compactified on $T^4\times T^2$\cite{Sen:2005pu}. Let $d(n)$ 
denote the degeneracy of heterotic string states carrying charge 
$N \mathbf{q}_e^2=2n$ -- the fractional charges arise from the 
twisted sectors in the CHL orbifolding. Every $1/2$-BPS 
multiplet/heterotic string state has degeneracy $16=2^{8/2}$. 
Then the generating function of $d(n)$ 
is\cite{David:2006ji,Dabholkar:2005by,Dabholkar:2005dt} 
\begin{equation}
\label{halfbps}
\sum_{n=0}^\infty d(n)\ \exp(\tfrac{2\pi i n \tau}N ) 
= \frac{16}{ (i\sqrt{N})^{-k-2}\ f^{(k)}(\tau/N)}\ ,
\end{equation}
where
\begin{displaymath}
f^{(k)}(\tau) \equiv  \eta(N\tau)^{k+2}\ \eta(\tau)^{k+2}\ .
\end{displaymath}  
The degeneracy of purely magnetically charged states with charge 
$\mathbf{q}_m=2 m$ is given by a similar formula with 
$f^{(k)}(\tau/N)$ replaced by $f^{(k)}(\tau)$. These are 
level-$N$ modular forms with weight $(k+2)$. For $(N,k)=(1,10)$, 
$f^{(10)}(\tau)=\eta(\tau)^{24}$.

\subsection{Counting $1/4$-BPS states}

As we saw earlier, $1/4$-BPS states are necessarily dyonic in 
character with the electric and magnetic charge vectors being 
linearly independent. In a remarkable leap of intuition, 
Dijkgraaf, Verlinde and Verlinde (DVV) proposed that a Siegel 
modular form of weight $10$ is the generating function of $1/4$-BPS 
states\cite{Dijkgraaf:1996it}. Let $d(n,\ell,m)$ be the 
degeneracy of $1/4$-BPS states with charges $\mathbf{q}_e^2=2n$, 
$\mathbf{q}_m^2=2m$ and $\mathbf{q}_e\cdot \mathbf{q}_m=\ell$. 
Then, one has
\begin{equation}
\label{DVVformula}
\sum_{(n,\ell,m)> 0} d(n,\ell,m)\ q^n\ r^\ell\ s^m =
\frac{64}{\Phi_{10}(\mathbf{Z})}\ ,
\end{equation}
with $\mathbf{Z}=\left(\begin{smallmatrix} z_1 & z_2 \\ z_2 & z_3 
\end{smallmatrix}\right) \in \BH_2$ and $q=\exp(2\pi i z_1)$, 
$r=\exp(2\pi i z_2)$, $s=\exp(2\pi i z_3)$ (see Appendix (B.1) 
for further details) and $(n,\ell,m)>0$ implies $n,m\geq1$, 
$\ell\in\BZ$ and $(4nm-\ell^2)>0$. The modular form 
$\Phi_{10}(\mathbf{Z})$ has a double zero at $z_2=0$, where it 
factorizes as
\begin{equation}
\label{factorise}
\lim_{z_2\rightarrow 0}\ \Phi_{10}(\mathbf{Z}) = (2\pi z_2)^2 \ 
\eta(z_1)^{24} \ \eta(z_3)^{24}
\end{equation}
In this limit, one sees the appearance of the degeneracies of the 
pure electric and magnetic states -- these are generated by 
$\eta(z_1)^{24}$ and $\eta(z_3)^{24}$ respectively. From this we 
see that the $\BZ_2$ transformation which exchanges electric and 
magnetic charges corresponds to the geometric action on $\BH_2$: 
$z_1\leftrightarrow z_3$. Another check of this formula is that 
this modular form (as well as the ones that we discuss later) 
matches the Bekenstein-Hawking entropy of large dyonic blackholes 
in the limit of large electric and magnetic charges\footnote{A 
more precise statement is that the formula reproduces the entropy 
of large blackholes with torsion one. The degeneracy of $1/2$-BPS 
blackholes is more complicated to understand as they have no 
horizon and hence have vanishing Bekenstein-Hawking entropy.}.

In a development that has spurred activity in this area, Jatkar 
and Sen generalized the DVV proposal to the case of asymmetric 
$\BZ_N$-orbifolds of the heterotic string on $T^6$ for 
$N=2,3,5,7$\cite{Jatkar:2005bh}.  They proposed that the 
degeneracy of $1/4$-BPS dyons is generated by a Siegel modular 
form of weight $k=\frac{24}{N+1}-2$ and level $N$, 
$\widetilde{\Phi}_k(\mathbf{Z})$. They also provided an explicit 
construction of the modular form using the additive lift of a 
weak Jacobi form. The constructed modular form has the following 
properties:
\begin{itemize}
\item[(i)] It is invariant under the S-duality group 
$\Gamma_1(N)$ suitably embedded in the group $G_1(N)\subset 
Sp(2,\BZ)$.
\item[(ii)] In the limit $z_2\rightarrow 0$, it has the right factorization
property:
\begin{equation}
\lim_{z_2\rightarrow 0}\ \widetilde{\Phi}_{k}(\mathbf{Z}) 
= (i\sqrt{N})^{-k-2}\ (2\pi z_2)^2 \
f^{(k)}(z_1/N) \ f^{(k)}( z_3)
\end{equation}
Note that for  $(N,k)=(1,10)$, this matches the DVV formula.
\item[(iii)] It reproduces the entropy for large
blackholes\cite{Jatkar:2005bh}.
\item[(iv)] For $1/2$-BPS blackholes, the formula leads to a 
prediction for $R^2$ (higher derivative) corrections to the 
low-energy effective action given in Eq. \eqref{lowenergy}. Such 
corrections lead to a non-zero entropy using Wald's 
generalization of the BH entropy formula for Einstein gravity 
that agrees with the prediction from the modular 
form\cite{Dabholkar:2004yr,Sen:2005iz}.
\end{itemize}
The S-duality group $\Gamma_1(N)$ as well as $\Gamma_0(N)\in 
SL(2,\BZ)$ are embedded into $Sp(2,\mathbb{Z})$ as follows\footnote{It
turns out that the modular forms constructed by Jatkar-Sen are invariant
under the larger group $\Gamma_0(N)$ for $N=2,3,5$.} :(Note 
that this is not quite the embedding given by Jatkar-Sen -- we 
have carried out a $z_1\leftrightarrow z_3$ exchange on their 
embedding to match our notation.)
\begin{equation}
\label{tildeembedding}
\left(\begin{array}{cc}a & b \\ c & d\end{array}\right) \mapsto
\left(\begin{array}{cccc}d & -c & c & 0 \\-b & a & 0 & b \\0 & 0 & a & b \\ 0 &
0 & c & d\end{array}\right)\ ,\quad c=0\mod N\ ,
\end{equation}
with the additional condition $a=1\mod N$ for $\Gamma_1(N)$.
Let us call this sub-group of $Sp(2,\mathbb{Z})$, $G_0(N)$. Further,
let $\Gamma^J(N)\equiv\Gamma^J \cap G_0(N)$. This is subgroup of $G_0(N)$ that
preserves the cusp at $(z_3)=i\infty$ (see appendix B). The weak Jacobi form
$\phi_{k,1}(z_1,z_2)$ that generates $\tilde{\Phi}_k(\mathbf{Z})$ is a modular
form of $\Gamma^J(N)$. Then, for  prime $N$, there are two inequivalent cusps in
the upper-half plane $\BH_1$ corresponding to $(z_1)=0$ and $(z_1)=i\infty$.
Thus, the modular form $\tilde{\Phi}_k(\mathbf{Z})$ behaves differently at the
two cusps. Its behavior at $z_1=0$ is captured by the modular form
$\Phi_k(\mathbf{Z})$ also defined by Jatkar and Sen\cite{Jatkar:2005bh}. 
One has
\begin{equation}
 \Phi_k(\mathbf{Z}) \equiv z_1^{-k} \
\widetilde{\Phi}_k(\mathbf{\widetilde{Z}})\ ,
\end{equation}
with
\[
\tilde{z}_1 = -1/z_1\quad,\quad \tilde{z}_2 = z_2/z_1\quad, \quad 
\tilde{z}_3 = z_3 -z_2^2/z_1\ .
\]
It is not hard to see that the above change of variables maps 
$(z_1)=0$ to $(z_1)=i\infty$ while preserving the cusp at 
$(z_3)=i\infty$. We will be dealing with the modular form, 
$\Phi_k(\mathbf{Z})$, for most of this paper as it is 
\textit{symmetric} in the `electric' and `magnetic' variables 
$z_1$ and $z_3$ even though it is $\widetilde{\Phi}_k(\mathbf{Z})$ 
which is the generating function of dyonic degeneracies. Note 
that this point is not relevant for $N=1$ (c.f. the DVV formula) 
as both the cusps get identified under $SL(2,\BZ)$.

\subsection{Product formulae for the modular forms}

Product representations for a closely related modular form 
$\Phi_k(\mathbf{Z})$ were provided by two 
groups\cite{David:2006ji,Dabholkar:2006xa}. In a subsequent 
paper\cite{David:2006yn}, David and Sen derived the product 
representations for modular form $\tilde{\Phi}_k(\mathbf{Z})$ as 
well as $\Phi_k(\mathbf{Z})$ using the 4D-5D correspondence with 
the well-studied $D1-D5$ system in Taub-NUT 
space\cite{Shih:2005uc}. They showed that it can be written as 
the product of three contributions arising from: (i) the 
low-energy dynamics in Taub-NUT space, (ii) the center of mass 
dynamics of the D1-D5 system in Taub-NUT space and (iii) the 
dynamics of D1-branes along $K3/\BZ_N$\footnote{The $\BZ_N$ is 
the Nikulin involution associated with the type IIA dual of the 
CHL orbifold.}. \begin{equation} \label{tildeproduct} 
\widetilde{\Phi}_k(\mathbf{Z})=(i\sqrt{N})^{-k-2}\ 
f_k(z_1/N)\times{\mathcal E}_{\textrm{st}\times 
T^2}(z_1,z_2)\times \hat{\mathcal{E}}_{S^*(K3/\BZ_N)} 
(z_1,z_2,z_3) \end{equation} where ${\mathcal 
E}_{\textrm{st}\times 
T^2}(z_1,z_2)=[\vartheta_1(z_1,z_2)/\eta(z_1)^3]^2$ is the 
spacetime elliptic genus\cite{Lerche:1999ju} and 
$\hat{\mathcal{E}}_{S^*(K3/\BZ_N)} (z_1,z_2,z_3)$ is the 
`second-quantized elliptic genus' of $K3/\BZ_N$ defined 
by\cite{David:2006yn,Dijkgraaf:1996xw} $$
 \hat{\mathcal{E}}_{S^*(K3/\BZ_N)}  (z_1,z_2,z_3)= \sum_{p=0}^{\infty}
\mathcal{E}_{S^p(K3/\BZ_N)} (z_1,z_2)\ e^{2\pi i p z_3}
$$
where $\mathcal{E}_{S^p(K3/\BZ_N)}$ is the elliptic genus of the 
$p$-th symmetric product of $K3/\BZ_N$ and 
$\mathcal{E}_{S^0(K3/\BZ_N)}\equiv1$. Note that each of the terms in 
Eq. \eqref{tildeproduct} independently admits a product 
representation -- for details see \cite{David:2006yn}.

We will obtain product representations for $\Phi_k(\mathbf{Z})$ 
for $N=2,3,5$ as the multiplicative (Borcherds) lift of a Jacobi 
form $\phi^{(N)}_{0,1}(\tau,z)$ of 
$\Gamma_0(N)^J=\Gamma_0(N)\ltimes \BZ^2$\cite{Aoki:2005}. For the 
cases of interest, the groups have two cusps at $\tau=i\infty$ 
and $\tau=0$. Let $c_1(n,\ell)$ and $c_2(n,\ell)$ be the 
coefficients of the Fourier expansion about the two cusps. Then, 
one has\cite{Aoki:2005,Dabholkar:2006xa}
\begin{equation}
\label{productformphik}
\Phi_k(\mathbf{Z}) = qrs \prod_{n,\ell,m\in \BZ} \Big(1-q^nr^\ell s^m\Big)^{c_1(nm,\ell)}\times 
\prod_{n,\ell,m\in \BZ} \Big(1-(q^nr^\ell s^m)^N\Big)^{c_2(nm,\ell)}\ .
\end{equation}
Note that we use a normalization for $\Phi_k(\mathbf{Z})$ that 
differs by on an overall sign from the one used by 
David-Jatkar-Sen\cite{David:2006ji}. To make use of this formula, 
we need to determine the relevant Jacobi form. According to 
Aoki-Ibukiyama (Prop. 6.1 in \cite{Aoki:2005}), any weak Jacobi 
form of weight zero and index one of $\Gamma_0(N)^J$ is of the 
form
\begin{equation}
\phi^{(N)}_{0,1}(\tau,z)= A^{(N)} \ \alpha^{(N)}(\tau)\ 
\phi_{-2,1}(\tau,z) + B^{(N)}\ \phi_{0,1}(\tau,z)\ ,
\end{equation}
where $\phi_{-2,1}(\tau,z)$ and $\phi_{0,1}(\tau,z)$ are as 
defined in Eq. \eqref{weakJacobiForm} and $\alpha^{(N)}(\tau)$ is 
a weight two modular form at level $N$ and $A^{(N)}$ and 
$B^{(N)}$ are constants. When $N=2,3,5$, there is only 
\textbf{one} such weight two modular form with constant 
coefficient $=1$ given by the Eisenstein series: 
$\alpha^{(N)}(\tau)=\tfrac{12i}{\pi(N-1)}\partial_\tau \big[\ln 
\eta(\tau) -\ln \eta(N\tau)\big]$. Thus, our problem is reduced 
to fixing two constant coefficients, $A^{(N)}$ and $B^{(N)}$ 
which we do now following the procedure given in ref. 
\cite{Dabholkar:2006xa}.

Recall that $\phi_{k,1}(\tau,z)$, that generates the additive 
lift for $\Phi_k(\mathbf{Z})$, provides all the terms in 
$\Phi_k(\mathbf{Z})$ that appear with coefficient 
$s$\cite{Jatkar:2005bh}. The product representation of the Jacobi 
form, $\phi_{k,1}(\tau,z)$, thus enables us to fix the 
coefficients in Eq. \eqref{productformphik} with $m=0$. Comparing 
the product form (referred to as the Hodge anomaly in 
\cite{Dabholkar:2006xa}) given below with Eq.  
\eqref{productformphik}
\begin{multline}
%\label{ }
\phi_{k,1}(\tau,z)=\eta(\tau)^{k-2}\eta(N\tau)^{k+2} \vartheta_1^2(\tau,z)\ , \\
=q r (1-r^{-1})^2 \prod_{n=1}^\infty (1-q^n)^{k-2} (1-q^{nN})^{k+2}(1-q^nr)^2(1-q^n r^{-1})^2\ ,
\end{multline}
we see that 
 \begin{equation}
 c_1(-1)=2\ , \quad c_1(0)=(k-2)\ ,\quad c_2(-1)=0\ \textrm{and}\quad c_2(0)=(k+2)\ ,
 \end{equation}
where the argument of $c_s$ is given in terms of $(4nm-\ell^2)$. 
Note that $(k+2)=24/(N+1)$ as usual. For instance, the 
coefficient of $r$ at the cusp at $i\infty$ implies $A ^{(N)} +(B 
^{(N)}/2)=c_1(-1)$ and similarly, the constant term at the same 
cusp implies $-2A ^{(N)} +5B ^{(N)} =c_1(0)$. This implies that
\begin{equation}
%\label{ }
A^{(N)}=\frac{2N}{N+1}\quad,\quad B^{(N)}=\frac1{N+1}\ ,
\end{equation}
which match the results given in ref. \cite{Dabholkar:2006xa} for 
$N=2$. These two conditions did not need too much detail about 
the weight two modular form $\alpha^{(N)}(\tau)$ other than its 
normalization, i.e., $\alpha^{(N)}(i\infty)=1$. The expansion at 
the other cusp provide an infinite number of consistency checks 
of the existence of the multiplicative lift.

The product expansion for $\Phi_k(\mathbf{Z})$ that has been 
given by David-Jatkar-Sen in \cite{David:2006ji} naively appears 
to be of a different form. However, one can show that the two 
expansions are indeed the same providing us with an additional 
check on our product formula.\footnote{We thank Justin David for 
useful discussions and in particular, for informing us about this 
relationship.} Further, it gives an alternate formula for 
$c_2(n,\ell)$ as the Fourier coefficients of another weak Jacobi 
form
\begin{equation}
\widehat{\phi}^{(N)}_{0,1}(\tau,z)= \frac{-2}{N+1} \ \alpha^{(N)}(\tau)\ \phi_{-2,1}(\tau,z) + \frac1{N+1}\ \phi_{0,1}(\tau,z)\ ,
\end{equation}
about the cusp at $i\infty$. It is also important, for our later 
discussion, to note that the coefficients $c_1(n,\ell)$ and 
$c_2(n,\ell)$ are all even integers for $N=2,3,5$ to the orders 
($n\leq 12$) that we have checked. We believe that this is true 
to all orders.

\section{Generalized Kac-Moody algebras}

\subsection{From Cartan Matrices to Lie Algebras} 

A Lie algebra is defined as a vector space $\Fg$ with an 
anti-symmetric bilinear map $[~,~] :\Fg \times \Fg \mapsto \Fg$ 
satisfying the Jacobi identity. A finite dimensional Lie algebra 
can also be defined through its Cartan matrix. Given a real, 
indecomposable, $(r \times r)$ symmetric matrix\footnote{The 
symmetric condition can be extended to include symmetrizable 
matrices. A matrix $A$ is said to be symmetrizable if there 
exists a non-degenerate diagonal matrix $D$ such that $A=DB$ 
where $B$ is a symmetric matrix.} $A = (a_{ij})$, $i, j \in I = 
\{ 1,2,\ldots,r\}$ of rank $r$ satisfying the following 
conditions:
\begin{itemize}
\item[(i)] $a_{ii} = 2$ ,
\item[(ii)]
$a_{ij} = 0 \Leftrightarrow a_{ji} = 0$ ,
\item[(iii)]$a_{ij} \in \mathbb{Z}_{\leq 0} \quad \textrm{for} 
\quad i \neq j$ ,
\item[(iv)]
$\textrm{det } A >0$\ ,
\end{itemize}
one defines a Lie algebra $\Fg(A)$ generated by the generators 
$e_i$, $f_i$, and $h_i$, for $i \in I$, satisfying the following 
conditions for $i,j \in I$ :
\begin{eqnarray}
[ h_i , h_j ] = 0\ ; \  [e_i,f_j]=\delta_{ij} h_j\ ;\ 
[ h_i , e_j ] = a_{ij} e_j\ ; \ [ h_i , f_j ] = - a_{ij} f_j\ ;
\quad \nonumber \\ 
(\textrm{ad }{e_i} )^{1-a_{ij}} e_j = (\textrm{ad }{f_i} )^{1-a_{ij}} f_j = 0 
\quad \textrm{ for } i \neq  j\ .\hspace{1in}
\nonumber
\end{eqnarray}
The Cartan matrix thus uniquely defines the Lie algebra which we call $\Fg(A)$
(and on occasion simply $A$).

More general classes of Lie algebras are obtained by weakening 
the axioms on the Cartan matrix. In particular, relaxing 
condition (iv) on the positive definiteness of the determinant of 
the Cartan matrix yields Kac-Moody algebras which are infinite 
dimensional. Relaxing conditions (i) and (iv), one obtains the 
most general class of Kac-Moody algebras, the generalized 
Kac-Moody (GKM) algebras.

Affine Lie algebras are obtained by requiring positive 
semi-definiteness of the determinant of $A$ in the place of 
positive definiteness. Thus, for an affine Lie algebra the 
condition on the determinant is,
\begin{equation} 
  \textrm{det}\  A_{\{i\}} >0 
\end{equation}
for all $i = 1,\ldots,r$, where $A_{\{i\}}$ are the matrices 
obtained from A by deleting the $i$-th row and column. For an 
affine Lie algebra, the rank of $A$ is at least $(r-1)$.

Our main focus will be on the following sequence of Cartan matrices
\begin{equation}\label{cartanmatrices}
A_1=\begin{pmatrix} 2 \end{pmatrix} \hookrightarrow
 A_1^{(1)}=\begin{pmatrix} ~~2 & -2 \\ -2 & ~~2 \end{pmatrix} \hookrightarrow
 A_{1,II}\equiv \begin{pmatrix} ~~2 & -2 &-2\\ -2 & ~~2&-2\\-2&-2&~~2 
\end{pmatrix}
\end{equation}
The first matrix in the sequence leads to the finite rank one Lie 
algebra $A_1$, the second leads to the affine Lie algebra 
$\hat{A}_1^{(1)}$. The Lie algebra $\Fg(A_{1,II})$ 
is a sub-algebra of the GKM superalgebras\footnote{A superalgebra 
is an algebra with a $\BZ_2$ grading -- the algebra has bosonic 
and fermionic elements satisfying a graded Lie bracket. The 
required generalizations have been discussed in 
\cite{Nikulin:1995} and we refer the reader to it.}, 
$\mathcal{G}_N$, that we construct in this paper.

\subsection{Cartan subalgebra}

The $\mathbb{Z}$-span $\Fh$, of the elements $\{h_i\}$ is called 
the \textit{Cartan subalgebra} of $\Fg(A)$.  The Lie algebra 
decomposes into eigenspaces, called root spaces, under the 
simultaneous adjoint action of $\Fh$. An element $ \alpha \in 
\Fh^{ \ast}$ is called a (real) root if the eigenspace
\begin{equation}
  \Fg_{\alpha} = \left\{ g \in \Fg ~|~ [ h, g] = \alpha (h)\ g, 
\forall  h \in \Fh\right\}
\end{equation}
defined by $\alpha$ is not empty. The set $L$, of all $\alpha$ 
such that $ \Fg_{\alpha} \neq 0$, is called the root system. The 
root space is generated by elements $\alpha_i$ satisfying
\begin{equation}
  \alpha_i ( h_j) = a_{ij} \quad \textrm{for}  \quad i, j  \in I.
\end{equation}
The elements $\alpha_i$ are called the simple roots and the set 
of roots are generated as integral linear combinations of these 
with coefficients either all positive, or all negative. These 
sets are called the set of positive and negative roots 
respectively, and are denoted by $L_+$ and $L_-$. Thus, $L = L_+ 
\cup L_-$. This decomposition gives a grading on the Lie algebra
\begin{equation}
\Fg = \bigoplus_{\alpha \in L} \Fg_{\alpha} = \bigoplus_{\alpha \in L_-}
\Fg_{\alpha} \bigoplus  \Fh \bigoplus_{\alpha \in L_+} \Fg_{\alpha} \ .
 \end{equation}

The affine Lie algebras are obtained by `adding' a root 
$\alpha_0$ to a finite Lie algebra. As the Cartan matrix of 
an affine Lie algebra is degenerate, there is an element, 
$k\in\Fh$, that is central, i.e., it commutes with all elements 
of the Lie algebra. For $\hat{A}_1^{(1)}$, one has $k=h_0+h_1$. 
The non-degeneracy of the Cartan matrix
is fixed by adding a new generator to $\Fh$ 
called the derivation, $d$, to the Lie algebra $\Fg(A_1^{(1)})$ 
with the following Lie brackets\cite{Kac:1990}:
\begin{equation}
[ \alpha_i , d ] = 0 ,\  [ \alpha_0 , d ] =+ \alpha_0,\ [ h_i, d] = 0,\  [k, d] = 0\ .
\end{equation}
The roots are defined on $\Fh^{\ast}$, consequently, modifying 
$\Fh$ implies the root space also gets modified accordingly. The 
set of positive roots, $L_+$, in this case are defined to be the 
union of the set of positive roots of the finite-dimensional Lie 
algebra and the set of roots for which
\begin{equation}
 [ \alpha , d ] = c\ \alpha\ , \textrm{ with the constant } c > 0\ .
 \end{equation}
For the affine Kac-Moody algebra, $\widehat{A}_1^{(1)}$, from the 
above definition of the set of positive roots, we have
\begin{equation}
L_+=\Big(n(\alpha_1+\alpha_0), n \alpha_1 + (n-1) \alpha_0, (n-1)\alpha_1+n 
\alpha_0~\Big|~ n=1,2,3,\ldots\Big)\ ,
\end{equation}
and $L_-=-L_+$.

The class of infinite dimensional algebras that are obtained from 
a generalized Cartan matrix are known as generalized Kac-Moody 
algebras have a more general root structure. In general there is 
no simple way to characterize the generalization from affine to 
the generalized class of Kac-Moody algebras, as there are many 
classes of GKM algebras that can be constructed due to the high 
degree of arbitrariness in the characterization of their Cartan 
matrices. The root systems are also considerably different from 
the affine case and can have imaginary simple roots in addition 
to real simple roots, and the multiplicity of the roots can be 
hard to determine.

The GKM algebra $\Fg(A_{1,II})$ is one which contains 
$\hat{A}_1^{(1)}$ as a sub-algebra and falls into a class of GKM 
algebras called hyperbolic Kac-Moody algebras that have been 
classified. There are infinitely many hyperbolic Kac-Moody algebras at 
rank two. At rank greater than two but less than eleven, there 
exist a finite number of hyperbolic Kac-Moody algebras. There are 
no hyperbolic Kac-Moody algebras at rank higher than ten. The 
embedding of $\hat{A}_1^{(1)}$ has been used by Feingold and 
Frenkel to study another rank three hyperbolic Kac-Moody 
algebra\cite{Feingold:1983}. More recently, Feingold and Nicolai 
have shown that this hyperbolic Kac-Moody algebra contains 
$\Fg(A_{1,II})$ as a sub-algebra\cite{Feingold:2003es}.

\subsection{Weyl groups}

The Weyl group of a Cartan matrix, $\mathcal{W}(A)$, is defined 
as the group generated by elementary reflections associated with the 
\textit{real} simple roots
\begin{equation}
w_{\alpha_i} (\beta) = \beta - 2 \frac{(\beta, \alpha_i)}{(\alpha_i, \alpha_i)} \alpha_i, 
\quad i \in I 
\end{equation}
and $\beta$ any root. It is easy to see that $w_{\alpha_i}^2=1$, 
$\forall i\in I$. Further, the Cartan matrix, $A$, determines 
further relations, if any, amongst the basic reflections and 
thus these groups are Coxeter groups. For symmetric matrices $A$, 
one has
\begin{equation}
\left(w_{\alpha_i} w_{\alpha_j}\right)^{2+|a_{ij}|^2}=1 
\textrm{ when } |a_{ij}|=0,1 
\textrm{ and } i\neq j\ .
\end{equation}
Further, when $|a_{ij}|\geq 2$, there are no relations. 

For the three Cartan matrices given in Eq. 
\eqref{cartanmatrices}, the Weyl group is generated by $r$ 
elementary reflections, $w_{\alpha_i}$ with no further relations. 
For $r=1$, this group is $\BZ_2$. For $r=2,3$, these groups are 
infinite dimensional. For $r=2$, the affine Weyl group is the 
semi-direct product $\left(\BZ_2\ltimes \BZ\right)$, where $\BZ$ 
is the normal sub-group of translations generated by $t\equiv 
\left(w_{\alpha_1} w_{\alpha_2}\right)$.

\subsubsection{The Weyl group $\mathcal{W}(A_{1,II})$}\label{weyldesc}

We now discuss the Weyl group associated with the Cartan matrix 
$A_{1,II}$ given in \eqref{cartanmatrices} in some detail as it 
plays an important role in this paper. Let 
$(\delta_1,\delta_2,\delta_3)$ be the three real simple roots 
whose Gram matrix (matrix of inner products) is given by the 
matrix $A_{1,II}$. Let $(w_1,w_2,w_3)$ denote the three 
reflections generated by the three simple roots. The Weyl group, 
$\mathcal{W}(A_{1,II})$, is generated by these three elementary 
reflections with no further relations. It turns out that 
$\mathcal{W}(A_{1,II})$ has a nicer presentation as a normal 
subgroup of $PGL(2,\BZ)$ which we describe later. One 
has\cite{Nikulin:1995} (see also 
\cite{Feingold:2003es,Cheng:2008fc})
\begin{equation}\label{isom}
PGL(2,\BZ) = \mathcal{W}(A_{1,II})\rtimes S_3\ ,
\end{equation}
where $S_3$ is the group of permutations of the three real simple roots.

The three real simple roots define the root lattice $M_{II} = 
\mathbb{Z} \delta_1 \oplus \mathbb{Z} \delta_2 \oplus \mathbb{Z} 
\delta_3 $ 
and a fundamental polyhedron, $\mathcal{M}_{II}$, which is given 
by the region bounded by the spaces orthogonal to the real simple roots. 
\begin{equation}
\mathbb{R}_+ \mathcal{M}_{II} = \{ x \in M_{II} \otimes \mathbb{R} ~|~ (x,
\delta_i) \leq 0, i = 1,2,3 \}.
\end{equation}
The lattice $M_{II}$ has a \textit{lattice Weyl vector} which is 
an element $\rho \in M_{II} \otimes \mathbb{Q}$ such that 
all the real simple roots satisfy\footnote{The standard convention is to
define $\rho$ through the condition
$(\rho,\delta_i)=(\delta_i,\delta_i)/2$ for all real simple roots
$\delta_i$. However, in this subsection, we reproduce the notation of
Gritsenko and Nikulin (which differs by a sign) here as we wish to compare 
with their results in the later part of the paper.}
\begin{equation}
 (\rho , \delta_i) = - \frac{(\delta_i , \delta_i)}2 =-1\  .
\end{equation}
One has $\rho = (\delta_1 + \delta_2 + \delta_3) /2$ i.e., 
it is one-half of the sum over real simple roots.
The positive real roots are then given by
\begin{equation}
L^{\textrm{re}}_+ = \Big(\mathcal{W}(\delta_1,\delta_2,\delta_3) 
\cap M^+_{II}\Big)\ ,
\end{equation}
where $\mathcal{W}$ refers to the Weyl group $\mathcal{W}(A_{1,II})$
and $M^+_{II} = \mathbb{Z}_+ \delta_1 \oplus 
\mathbb{Z}_+ \delta_2 \oplus \mathbb{Z}_+ \delta_3$.

In order to make the map to $PGL(2,\BZ)$ explicit, we will choose 
to write the roots in terms of a basis $(f_2, f_3,f_{-2})$ 
which are related to the $\delta_i$ in the 
following way:
\begin{equation}
\delta_1 = 2 f_2 - f_3,\quad\delta_2 = f_3 ,\quad \delta_3 = 2 f_{-2} - f_3  \ .
\end{equation}
The non-vanishing inner products among the elements $f_i$  are: 
\begin{equation}
  (f_2, f_{-2}) = -1 ,\quad (f_3, f_3) = 2\ .
\end{equation}
Thus, $(f_2,f_3,f_{-2})$ provide a basis for Minkowski space $\BR^{2,1}$. 
Consider the time-like region
\[
 V = \{ x\in \BR^{2,1}~|~ (x,x) <0\} \ , 
\]
in $\BR^{2,1}$. Let $V^+$ denote the future light-cone in the 
space and
\begin{equation}
\mathbf{Z}=z_3 f_2 + z_2 f_3 + z_1 f_{-2}\ ,
\end{equation}
be such that $\mathbf{Z}\in \BR^{2,1}+ i V^+$. This is equivalent 
to $\mathbf{Z}\in \BH_2$, the Siegel 
upper-half space\cite{Nikulin:1995}.
 
Let $M_{1,0}$ be the lattice $(\BZ f_2 + \BZ f_3 + \BZ f_{-2})$. 
The root lattice $M_{II}$ is a sub-lattice in $M_{1,0}$. 
$M_{1,0}$ happens to be the root lattice of another rank three 
hyperbolic Kac-Moody algebra\cite{Feingold:1983,Feingold:2003es}. 
The Weyl group of this algebra is isomorphic to $PGL(2,\BZ)$. 
Consider the following identification:
\begin{equation}
%\label{ }
f_{-2} \leftrightarrow \begin{pmatrix}1 & 0 \\0 & 0\end{pmatrix} \ , \
f_{3}\leftrightarrow \begin{pmatrix}0 & 1 \\1 & 0\end{pmatrix}\ ,\
f_{2} \leftrightarrow \begin{pmatrix}0 & 0 \\0 & 1\end{pmatrix}\ .
\end{equation}
The norm of a matrix $N \in M_{1,0}$ is then given by $-2 \det 
N$. The Weyl group has the following action:
\begin{equation}
%\label{ }
N \rightarrow A \cdot N \cdot  A^T\ , \quad A \in PGL(2,\BZ) \textrm{ and
} N \in M_{1,0}\ .
\end{equation}
Recall that $PGL(2,\BZ)$ is given by the integral matrices 
$\left(\begin{smallmatrix} a & b \\ c & d \end{smallmatrix} 
\right)$ with $ad-bc=\pm 1$. The $S_3$ mentioned 
in Eq. \eqref{isom} is generated by
\begin{equation}
%\label{ }
r_{-1}=\begin{pmatrix} 0 & 1 \\ 1 & 0 \end{pmatrix}\quad,\quad
r_0 =\begin{pmatrix} 1 & 1 \\ \ 0 & -1 \end{pmatrix}\ .
\end{equation}
satisfying $r_{-1}^2=r_0^2 = (r_{-1}r_0)^3=1$. The three 
elementary reflections that generate $\mathcal{W}(A_{1,II})$ are 
given by the following $PGL(2,\BZ)$ matrices:
\begin{equation}
w_{\delta_1}= \begin{pmatrix} -1 & 0
\\ 2 & 1 \end{pmatrix}\ , \
w_{\delta_2} = \begin{pmatrix} 1 & 0 \\ 0 & -1 \end{pmatrix}\ , \
w_{\delta_3}=\begin{pmatrix} 1 & 2\\ 0 & -1 \end{pmatrix}\ .
\end{equation}

\subsection{The Weyl denominator formula} 
\subsubsection{Finite Lie algebras}

The Weyl denominator formula is a specialisation of the Weyl 
character formula for the trivial representation. For finite Lie 
algebras, one has\footnote{Conventionally, the denominator 
formula is written with $e^\rho$ multiplying the form given here. 
We write in a form that is more suitable to affine Kac-Moody 
algebras.}
\begin{equation}
\prod_{\alpha\in L_+} \ \left(1 -e^{-\alpha}\right) = \sum_{w\in \mathcal{W}} 
\det (w)\ e^{[w(\rho)-\rho]}\ ,
\end{equation}
where $e^\alpha$ is a formal exponential and the Weyl vector 
$\rho$ is defined to be half the sum of all positive roots, i.e., 
$\rho=\tfrac12 \sum_{\alpha\in L_+} \alpha$. Further, $w(\rho)$ 
is the image of $\rho$ under the action of the element $w$ of the 
Weyl group.

\subsubsection{Affine Kac-Moody algebras}

The first twist occurs for the affine Kac-Moody algebras, where 
one needs to include imaginary roots, specifically those with 
zero norm, into the set of positive roots. Thus, one has 
$L_+=L_+^{\textrm{re}}\cup L_+^{\textrm{im}}$ and so on. Further, 
the imaginary roots appear with multiplicity not necessarily equal to one. 
For instance, for non-twisted affine algebras, the multiplicity of imaginary roots is equal to  the rank of the  underlying finite dimensional Lie algebra.

Also, the definition of the Weyl vector as half the sum over all 
positive roots, needs to be regulated due to the infinite number 
of such positive roots.\footnote{There is another definition of the Weyl
vector $\rho$, through its inner product with all real simple roots
$\delta_i$. It is the vector that satisfies
$(\rho,\delta_i)=\tfrac12(\delta_i,\delta_i)$, $\forall i$.} 
An alternate definition is given by 
Lepowsky and Milne which is tailored to writing the sum side of 
the denominator formula\cite{Lepowsky:1978}. The key observation 
(due to MacDonald\cite{MacDonald:1972}) is that $[w(\rho)-\rho]$ 
behaves better than 
either of the terms. Recall that an element of the Weyl group 
acts as a permutation of all roots (not necessarily positive). 
Thus, $[w(\rho)-\rho]$ obtains contribution, only when a positive 
root gets mapped to a non-positive root. So one defines the set 
$\Phi_w$ for all $w\in \mathcal{W}$, \begin{equation} \Phi_w 
=w(L_-)\cap L_+ =\Big\{ \alpha\in L_+ ~\Big|~ w^{-1}(\alpha)\in 
L_-\Big\}\ . \end{equation} Using this definition, we can see 
that \begin{equation} \rho-w(\rho)=\frac12 \sum_{\alpha\in L_+} 
\left[\alpha-w(\alpha)\right] \sim \langle \Phi_w \rangle\ , 
\end{equation} where $\langle \Phi_w \rangle$ is the sum of 
elements of the set $\Phi_w$. Note that
$-L_-=L_+$, which explains the half disappearing in the RHS of 
the above formula. Imaginary roots do not appear in the set 
$\Phi_w$ for affine Lie algebras as the imaginary roots turn out 
to be Weyl invariant and hence cancel out in the  above 
equation.
 
The denominator formula that works for affine Kac-Moody algebras, 
after including the imaginary roots in $L_+$, is the 
Weyl-Kac denominator formula
\begin{equation}
\prod_{\alpha\in 
L_+} \ \left(1 -e^{-\alpha}\right)^{\textrm{mult}(\mathbf{\alpha})} = 
\sum_{w\in \mathcal{W}} \det (w)\ e^{-\langle\Phi_w\rangle}\ 
,
\end{equation}

For the affine Kac-Moody algebra, $\widehat{A}_1^{(1)}$, from the 
above definition of the set of positive roots, we have
\begin{equation}
L_+=\Big(n(\alpha_1+\alpha_0), n \alpha_1 + (n-1) \alpha_0, 
(n-1)\alpha_1+n \alpha_0~\Big|~ n=1,2,3,\ldots\Big)\ ,
\end{equation}
and the Weyl group is isomorphic to $\BZ_2\ltimes \BZ$. Putting it all together
into the denominator identity gives\cite{Lepowsky:1978}
\begin{multline}
 \prod_{n \geq 1} ( 1 - e^{-n \alpha_0} e^{-n\alpha_1}) 
( 1 - e^{-(n-1) \alpha_0} e^{-n\alpha_1}) ( 1 - e^{-n \alpha_0} 
e^{-(n-1)\alpha_1}) \\
= \sum_{n \in \BZ} e^{-n(2n-1)\alpha_0} e^{-n(2n+1)\alpha_1} - 
\sum_{n \in \BZ} e^{-(n+1)(2n+1)\alpha_0} e^{-n(2n+1)\alpha_1}
\end{multline}
Setting $e^{-\alpha_0}=r$ and $e^{-\alpha_1}=qr^{-1}$, the above identity is
equivalent to the Jacobi triple identity involving the  theta function
$\vartheta_1(\tau,z)$:
\begin{eqnarray}
-i\vartheta_1(\tau,z)&=& q^{1/8} r^{-1/2} \prod_{n=1}^\infty \
\left(1-q^n\right)\
\left(1-q^{n-1}r\right)\ \left(1-q^{n} r^{-1}\right) \nonumber \\
&=&  \sum_{n\in {\mathbb Z}}\ (-1)^n\ q^{\frac{(n-1/2)^2}2}\ r^{n-1/2}\ .
\end{eqnarray}

\subsubsection{Generalized Kac-Moody algebras}

For the case of generalized Kac-Moody algebras, the story gets a 
bit more involved. A detailed discussion of the Weyl denominator 
formula for GKM algebras can be found in \cite{Lepowsky:1976}. 
New kinds of roots appear here and have nontrivial multiplicities 
that are hard to determine. The `sum side' denominator formula 
also requires modification. For the class of GKM algebras with a 
bilinear form that is almost positive definite the denominator 
identity was constructed by Borcherds generalizing the Weyl-Kac 
character formula for Kac-Moody 
algebras\cite{Borcherds:1988pq,Borcherds:1990,Borcherds:1992,Jurisich:1996}. 
These GKM algebras contain, in addition to real and imaginary 
roots of Kac-Moody algebras, imaginary simple roots and a 
suitable definition of the denominator identity is required to 
correct for these. Also, the way the Cartan subalgebra is 
centrally extended modifies the root space and requires some care 
as the linear independence of the simple roots is crucial for the 
denominator identity to be well defined\cite{Jurisich:1996}. The 
Weyl group continues to be generated only by elementary 
reflections of the \textit{real} simple roots.

For the above class of generalized Kac-Moody algebras, the 
Weyl-Kac-Borcherds denominator identity 
becomes\cite{Borcherds:1988pq}
\begin{equation}
\label{WKBformula}
\prod_{\alpha \in L_+} (1- e^{- \alpha})^{\textrm{mult}(\alpha)} = e^{-\rho}\
\sum_{w \in \mathcal{W}} (\textrm{det} w)\  w(e^{\rho} \sum_{\alpha \in L_+}
\epsilon (\alpha) e^{\alpha} )\  , 
\end{equation}
where $L_+$ is the set of positive roots, $\rho$ the Weyl vector, 
$\mathcal{W}$ the Weyl group of the GKM, det$(w)$ is defined to 
be $\pm1$ depending on whether $w$ is the product of an even or 
odd number of reflections and $\epsilon (\alpha)$ is defined to 
be $(-1)^{n}$ if $\alpha$ is the sum of $n$ pairwise independent, 
orthogonal imaginary simple roots, and 0 otherwise. As we will
consider superalgebras, the above formula, Eq. \eqref{WKBformula} 
continues to hold with roots appearing with graded multiplicity 
-- fermionic roots appear with negative multiplicity.

As before we can cast it into a slightly different form by 
constructing the set
\begin{equation}
\Phi_w =w(L_-)\cap L_+ =\Big\{ 
\alpha\in L_+ ~\Big|~ w^{-1}(\alpha)\in L_-\Big\}\ .
\end{equation}
such that 
\[ 
\rho-w(\rho)= \langle \Phi_w \rangle\ 
\]
and the denominator identity takes the form\cite{Jurisich:1996}
\begin{equation}
\prod_{\alpha \in L_+} (1- e^{- \alpha})^{\textrm{mult}(\alpha)} 
= \sum_{w \in W^{\sigma}} (\textrm{det} w) \sum_{\eta \in \Omega} 
(-1)^{\textrm{ht} (\eta)} e^{-\langle \Phi_w \rangle\ - w(\eta) }\ , 
\end{equation}
where $\Omega$ is defined as the sum of all the possible 
sets of distinct pairwise orthogonal imaginary simple roots and the 
height $\textrm{ht} (\eta)$ of an element $\eta = \sum_{i} 
n_i \delta_i$ is $\sum_{i} n_i$.
We will verify that the modular forms that we construct have product 
and sum representations that are indeed compatible with the 
denominator formula.

\section{Denominator formulae for $\Delta_{k/2}(\mathbf{Z})$}

Gritsenko and Nikulin have shown that the denominator formula of a 
GKM superalgebra $\mathcal{G}_1$ is related to a modular form of 
$\textrm{Sp}(2,\BZ)$ with character, $\Delta_5(\mathbf{Z})$.
The modular form, $\Phi_{10}(\mathbf{Z})$, 
that generates the degeneracy of dyons is equal to 
$\Delta_5(\mathbf{Z})^2$. We extend this correspondence to argue 
for the existence of new GKM superalgebras, $\mathcal{G}_N$, 
whose denominator formulae give rise 
to modular forms with character, $\Delta_{k/2}(\mathbf{Z})$, of 
suitable groups, $G_0(N)\subset \textrm{Sp}(2,\BZ)$, with the 
property
\begin{equation}
\label{Deltadef }
\Delta_{k/2}(\mathbf{Z})^2 = \Phi_k(\mathbf{Z})\ . 
\end{equation}
(Note that the $\Phi_k(\mathbf{Z})$ used here differs from the 
one defined by David-Jatkar-Sen\cite{David:2006ji} by an overall 
sign.) This is the \textit{main} assumption of this paper. Recall 
that the modular forms $\Phi_k(\mathbf{Z})$ were first 
constructed in \cite{Jatkar:2005bh} and product formulae were 
provided in subsequent 
papers\cite{David:2006ji,Dabholkar:2006xa}. Further, the dyon 
degeneracies are given by a closely related modular form called 
$\tilde{\Phi}_k(\mathbf{Z})$ by Jatkar and Sen. This differs in 
the way the S-duality group $\Gamma_1(N)\subset SL(2,\BZ)$ is 
embedded in $Sp(2,\BZ)$.

\subsection{Constructing the modular form $\Delta_{k/2}(\mathbf{Z})$}

We have found experimentally that the modular form 
$\Delta_{k/2}(\mathbf{Z})$ can be obtained as the additive lift 
of the Jacobi cusp form of $\Gamma_1(N)$ of weight $k/2$ and 
index $1/2$
\begin{equation}
\label{AddJacobiForm}
\psi_{k/2,1/2}(\tau,z) = \theta_{1}(\tau,z)\ \eta(\tau)^{(k-4)/2}\ 
\eta(N \tau)^{(k+2)/2}\ .
\end{equation}
A proof of the additive lift has been given in appendix \ref{modularity}.

Note that in the following we will write $z_1$ for $\tau$ and 
$z_2$ for $z$ keeping in mind that these become two of the three 
coordinates on $\BH_2$. This happens to be the square root of the 
Jacobi form that generates $\Phi_k(\mathbf{Z})$. A discussion of 
the additive lift has been provided in appendix B. The Fourier 
expansion
\begin{equation}
\psi_{k/2,1/2}(z_1,z_2)=\sum_{n,\ell\equiv 1 \textrm{ mod } 2} g(n,\ell)\ 
q^{n/2}  r^{\ell/2}\ ,
\end{equation}
with $q=\exp(2\pi i z_1)=\exp(2\pi i \tau)$ and $r=\exp(2\pi i 
z_2)=\exp(2\pi i z)$, enables us to define the additive lift for 
$\Delta_{k/2}(\mathbf{Z})$:
\begin{equation}
\label{DeltaMaass}
\Delta_{k/2}(\mathbf{Z}) \equiv \sum_{(n,\ell,m)>0}\ \  
\sum_{d | (n,\ell,m)} \ \chi(d)\  
d^{\frac{k-2}2}\ g\left(\tfrac{nm}{d^2},\tfrac\ell{d}\right)\ 
q^{n/2} r^{\ell/2} s^{m/2}\ ,
\end{equation}
where $s=\exp(2\pi i z_3)$. This generalises the Maa\ss\ formula 
for $\Delta_5(\mathbf{Z})$ when $(N,k)=(1,10)$. We have 
experimentally verified that the square of the above formula 
generates $\Phi_k(\mathbf{Z})$ to fairly high order (all terms 
which appear with  values of $mn\leq 15$ in the Fourier 
expansion for $\Phi_k(\mathbf{Z})$ and more to confirm the 
non-trivial character for $N=3$). We find that for $N=2,5$, the 
character $\chi(d)$ is the trivial one (see Eq. 
\eqref{trivialchar}) while for $N=3$, we need a non-trivial 
character $\chi^\psi(d)=\left(\frac{-3}{d}\right)$ i.e.,
\begin{equation}\label{modthree}
\chi^\psi(d)=\left\{\begin{array}{ll}\phantom{-}0&  d=0\mod 3\ , \\ 
\phantom{-}1 & d=1\mod 3\ ,\\ -1 & d=2 \mod 3\ . \end{array}\right.
\end{equation}

It is known that $\Delta_5(\mathbf{Z})$ can be written as a 
product of all ten even genus-two theta constants.  As we show in 
the appendix, $\Delta_3(\mathbf{Z})$ can be written as a product 
of six even genus-two theta constants:
\begin{equation}
\Delta_3(\mathbf{Z}) = \frac1{64}\ \theta_2(\mathbf{Z})\! 
\prod_{m=1\textrm{ mod }2} \theta_m(\mathbf{Z})\ .
\end{equation}
We however do not expect such a formula for the other 
$\Delta_{k/2}(\mathbf{Z})$.

It is of interest to ask whether these modular forms already 
exist in the mathematical literature. For $N=2$, we have been 
able to show that $\Delta_3(\mathbf{Z})$ is given by the product 
of six even genus-two theta constants. In fact, the square root 
of $\tilde{\Phi}_6(\mathbf{Z})$, which we call 
$\tilde{\Delta}_3(\mathbf{Z})$, can also be written as the 
product of six (different) even genus-two theta constants (see 
appendix A.2 and ref. \cite{Aoki:2005,Dabholkar:2006bj}). Aoki 
and Ibukiyama\cite{Aoki:2005} have studied the ring of modular 
forms at levels $N=2,3,4$. At level $N=3$, they show that modular 
forms with character, 
\[v^\psi(M)=\left(\frac{-3}{\det(D)}\right)\] for 
$M=\left(\begin{smallmatrix} A & B \\ C & 
D\end{smallmatrix}\right)\in \widehat{G}_0(3)$, necessarily have 
odd weight. However, for $N=3$, we obtain a modular form with 
even weight, $\Delta_2(\mathbf{Z})$. \textit{Is there a 
contradiction?} The resolution lies in the fact that 
$\Delta_2(\mathbf{Z})$ transforms with a different 
character\footnote{We thank Prof. Aoki for encouraging us to 
believe in the existence of $\Delta_2(\mathbf{Z})$ even though 
our original approach initially had no mathematical rigor. He has 
also informed us that Gritsenko has independently constructed a 
modular form of weight two at level 3. We also thank D. 
Ramakrishnan for a useful conversation.} -- it is the one given 
by the product of the character, $v^\psi(M)$, considered by 
Aoki-Ibukiyama and the restriction of the character, 
$v^\Gamma(M)$, that appears in the transformation of 
$\Delta_5(\mathbf{Z})$, to $\widehat{G}_0(3)$(see appendix B.4). 
In appendix \ref{genusone}, we show that the Jacobi form that 
generates $\Delta_2(\mathbf{Z})$ is one with non-trivial 
character which is consistent with our expectations. This 
combined with the additive lift shown in appendix 
\ref{modularity} show that this indeed true.

With the construction of $\Delta_{k/2}(\mathbf{Z})$, 
we are ready to verify whether they can be 
the denominator formula of a GKM superalgebra $\mathcal{G}_N$. 
The product formula for 
$\Delta_{k/2}(\mathbf{Z})$ can be obtained from Eq. 
\eqref{productformphik} and it gives us the positive roots $L_+$ with 
their multiplicities. 
\begin{equation}
\label{productformdeltak}
\Delta_{k/2}(\mathbf{Z}) = q^{1/2}r^{1/2}s^{1/2} \prod_{n,\ell,m\in \BZ} 
\Big(1-q^nr^\ell s^m\Big)^{\tfrac12 c_1(nm,\ell)}\times 
\prod_{n,\ell,m\in \BZ} \Big(1-(q^nr^\ell s^m)^N\Big)^{\tfrac12 c_2(nm,\ell)}\ .
\end{equation}
For this to make sense as the product side of a denominator 
formula of a GKM superalgebra, it is necessary that 
$c_1(nm,\ell)$ and $c_2(nm,\ell)$ are \textit{even} integers as 
they provide the multplicity of various roots in $L_+$. It 
appears to be true to the orders that we have checked. The common 
factor $q^{1/2}r^{1/2}s^{1/2}$ can be identified with exp$(- \pi 
\imath (\rho,z) )$ giving us the Weyl vector $\rho$.  We thus see 
that the Weyl vector is independent of $N$. The $\rho$ vector 
does not change during the orbifolding process in the work of 
Niemann either (see Theorem (1.6) in ref. \cite{Niemann}). In the 
work of Niemann, the orbifolding changes the simple roots whereas 
in our situation, the three real simple roots are unaffected by 
the orbifolding.

However, one needs to verify that the Maa\ss\ formula 
\eqref{DeltaMaass} corresponds to the sum side of the denominator 
formula. This is the procedure adopted by Gritsenko and Nikulin 
for $\Delta_5(\mathbf{Z})$ and we repeat their 
method\cite{Nikulin:1995}.

\subsection{The GKM superalgebra $\mathcal{G}_1$}

We first consider the example of $\mathcal{G}_1$ to illustrate 
the method before moving on to the new GKM superalgebras. As it 
stands, the denominator formula for the Lie algebra 
$\Fg(A_{1,II})$ is not related to the the automorphic form 
$\Delta_5(\mathbf{Z})$. Following the ideas of 
Borcherds\cite{Borcherds:1988pq,Borcherds:1990}, Gritsenko and 
Nikulin constructed a superalgebra, $\mathcal{G}_1$ by adding 
imaginary simple roots -- some are bosonic and others are 
fermionic.  The superalgebra $\mathcal{G}_1$ has $\Fg(A_{1,II})$ 
as a sub-algebra and its Weyl group is identical to the one for 
$\Fg(A_{1,II})$.

In the above setting, the Weyl-Kac-Borcherds denominator 
formula\footnote{Here we write the denominator formula in the 
notation of Gritsenko and Nikulin. In particular, one needs to 
replace $\rho$ by $-\rho$ in Eq. \eqref{WKBformula}.} 
\eqref{WKBformula} becomes (see also sec. \ref{weyldesc})
\begin{multline}
\label{match}
e^{- \pi \imath (\rho, z)}\prod_{\alpha\in L_+}\left(1-e^{-\pi\imath(\alpha,z)}\right)^{\textrm{mult}(\alpha)}  \\
=
\left(
\sum_{w \in \mathcal{W}} \textrm{det} (w) \left\{ e^{- \pi \imath (w (\rho),
z)} - \sum_{\eta \in M_{II} \cap \mathbb{R}_{+} \mathcal{M}_{II}} \! \!\!\!
m(\eta)\ e^{- \pi \imath (w(\rho + \eta ) , z)}\right\}\right) 
\end{multline}
where the element $\mathbf{Z} = z_3 f_2 + z_2 f_3 + z_1 f_{-2}$ 
belongs to the subspace $\BR^{2,1} + \imath V^+ 
\sim\BH_2$ obtained upon complexification of the cone 
$V^+$. We have deliberately rewritten the sum as two 
terms -- one arising from the real simple roots ($\eta=0$) and 
the other arising from the imaginary simple roots ($\eta\neq 0$). 
The first term thus arises as the sum side of the Lie 
algebra $\Fg(A_{1,II})$. The second terms is specific to GKM 
algebras due to the presence of imaginary simple roots with 
`multiplicities' $m(\eta) \in \mathbb{Z}$. These 
multiplicities are determined by the connection with the 
automorphic form $\Delta_5(\mathbf{Z})$(an explicit expansion is 
given in appendix D) -- in other words, one adds enough imaginary 
simple roots such that the automorphic properties are attained.  
The imaginary simple roots belong to the space $M_{II} \cap 
\mathbb{R}_{+} \mathcal{M}_{II}$. The Maa\ss\ formula for 
$\Delta_5(\mathbf{Z})$ leads to a simple expression for 
light-like imaginary simple roots i.e., $(\eta_0, \eta_0) = 0$. 
They are generated by the formula
\begin{equation}
1- \sum_{t \in \mathbb{N}} m(t \eta_0) q^t 
= \prod_{k \in \mathbb{N}} (1-q^k )^9
= \frac{\sqrt{ f^{(10)}(\tau)}}{\eta(\tau)^3}\  .
\end{equation}
Negative values of multiplicity implies that the root is 
fermionic. For instance, one has $m(2 \eta_0)=-27$. Thus, such 
roots are fermionic and hence we have a superalgebra. There are 
three primitive light-like vectors: $2 f_2$, $2 f_{-2}$ and $(2 
f_{-2} - 2f_3 + 2f_{2})$ each with multiplicity $9$. The action 
of the Weyl group generates the remaining vectors. There are two 
primitive vectors satisfying $(\eta, \eta) < 0$ : $(2 f_{-2} + 
2f_{2})$ and $(2 f_{-2} - f_3 + 2f_{2})$. The other terms are 
generated as multiples of the form $t \eta$ of these primitive 
vectors and multiplicities given by the above formula.
 
The product formula for $\Delta_5(\mathbf{Z})$ determines the 
positive roots $L_+$ -- again fermionic roots appear with 
negative multiplicity in the exponent. We do not write out the 
detailed list and refer the diligent reader to the paper by 
Gritsenko and Nikulin\cite{Nikulin:1995}. We would like to 
comment here that there is a subtle issue in
extracting the multiplicities from the exponent in the product 
formula -- the product formula  gives only the 
difference between the multiplicities of the bosonic and 
fermionic generators and hence is more like a Witten index.

\subsection{The GKM superalgebra $\mathcal{G}_2$}

An important assumption in this paper has been that the square 
root of the Siegel modular forms $\Phi_k(\mathbf{Z})$, i.e., 
$\Delta_{k/2}(\mathbf{Z})$ should appear as the denominator 
formula of some GKM superalgebra. The first example that we use to
verify this is the $\BZ_2$ orbifold for which $k=6$. It might be 
thought that although $\Phi_{10}(\mathbf{Z})$ and 
$\Phi_6(\mathbf{Z})$ are not obviously associated to any GKM 
algebras, we could nevertheless obtain the changes that result 
from the orbifolding by directly comparing the two as given by 
their Maa\ss\ expansions. The remarkable thing is that the 
comparison at the level of the modular forms, 
$\Phi_{10}(\mathbf{Z})$, and $\Phi_6(\mathbf{Z})$, is not very 
transparent, in that the terms appearing in the two are not one-to-one 
for comparison. There occur terms in each that do not appear in the 
other, and hence we can not track down their fate as we orbifold. 
It is only when we compare $\Delta_5(\mathbf{Z})$ and 
$\Delta_3(\mathbf{Z})$ we see that both the expansions are similar
and suitable for comparison. Below we analyze the Maa\ss\ 
expansion of its square root, $\Delta_3(\mathbf{Z})$, in the same 
way as was done for $\Delta_5(\mathbf{Z})$ and obtain the 
multiplicities of the primitive imaginary simple roots $\eta$.

We have already derived a Maa\ss\ formula for 
$\Delta_3(\mathbf{Z})$ using the additive lift of the weak Jacobi 
form $\psi_{3,1/2}(z_1,z_2)$ of weight $3$ and index $1/2$ as discussed 
earlier.  We also need to 
suitably interpret the region of summation of the variables 
$n,\ell,m$ such that they give the space $\Omega$ of the algebra 
we construct. Mathematically, these will be the `twisted' 
generalized GKM algebras that are obtained by the orbifold 
action on $\mathcal{G}_1$ analogous to the ones constructed by 
Niemann\cite{Niemann}.
 
A straightforward comparison of the Fourier expansions 
$\Delta_3(\mathbf{Z})$ and $\Delta_5(\mathbf{Z})$ (given in 
Appendix D) by first focusing on terms appearing from real simple 
roots, we observe that the Cartan matrix for $\mathcal{G}_2$ is 
the same as for $\mathcal{G}_1$. Next, we can see that all terms 
that appear in $\Delta_5(\mathbf{Z})$ appear in 
$\Delta_3(\mathbf{Z})$ albeit with different coefficients. Thus 
the set of simple roots -- real and imaginary, of $\mathcal{G}_2$ 
remain unchanged from $\mathcal{G}_1$. This implies the lattice 
generated by the three real simple roots \textit{$M_{II}$} = 
$\mathbb{Z} \delta_1 \oplus \mathbb{Z} \delta_2 \oplus \mathbb{Z} 
\delta_3 $ is the same as before. It also implies the cone 
defined by the lattice $V^+ (M_{II})$ and the fundamental 
polyhedron $\mathcal{M}_{II}$, generated by the spaces bounded by 
the real roots, also remain unchanged. The fundamental cone, 
$M_{II} \cap \mathbb{R}_{+} \mathcal{M}_{II}$, containing the the 
imaginary roots also remains unchanged by orbifolding. As already 
observed, the Weyl vector for the lattice $\rho$ is the same as 
for $\mathcal{G}_1$.

The only difference comes in the generating function of the 
multiplicity factors such as $m(\eta_0)$. The $m(\eta_0)$ in the 
case of $\mathcal{G}_2$ are given by the generating function
\begin{equation}
1 - \sum_{t \in \mathbb{N}} m(t \eta_0)\ q^t = \prod_{k \in \mathbb{N}} (1 -
q^k) ( 1- q^{2k})^4= \frac{\sqrt{f^{(6)}(\tau)}}{\eta(\tau)^3} \
\end{equation}
This can be understood from the twisted denominator formula of 
Niemann\cite{Niemann} for the cycle shape $1^{k+2}N^{k+2}$ for $N 
= 2$, and $k = 6$ and is the special case of a more general 
formula for any $N$ and $k$ which we give below.

It is remarkable that although the degeneracy of the dyons are 
given by completely different modular forms in the two cases, the 
underlying GKM that we can construct for them are similar to 
such a degree. The roots -- real and imaginary, Weyl group, Weyl 
vector, the lattice, and the space of the imaginary roots remain 
unchanged. The orbifolding of the space is reflected only in the 
change of the multiplicity of the imaginary roots.

\subsection{The GKM superalgebras $\mathcal{G}_N$}

In this section we list the properties of the class of GKM 
algebras $\mathcal{G}_N$ for $N = 1,2,3,5$ obtained from the 
modular forms that occur in degeneracy formulae for a 
$\mathbb{Z}_N$ orbifolding action.
 
The Cartan matrix, Weyl group, and the set of real and imaginary 
simple roots for the $\mathcal{G}_N$ remain the same for all 
values of $N$. The modular forms leading to these algebras, and 
hence the denominator identities of the algebras, however, are 
different from each other. The difference in the denominator 
identities is in the coefficients of the terms occurring in the 
expansion, whereas the terms themselves undergo no change.

The generating functions of the multiplicity factors of 
light-like simple roots for different values of $N$ are given in 
terms of a single formula\footnote{Interestingly, it was this 
pattern that lead us to the proposed additive lift using the weak 
Jacobi form $\psi_{k/2,1/2}(z_1,z_2)$.}:
\begin{equation}
\theta_1(\tau,z)\ \left( 1 - \sum_{t \in \mathbb{N}}
 m(t \eta_0)\ q^t\right) = \psi_{k/2,1/2}(\tau,z)\ .
\end{equation}
From the above we see the pattern in the progression of the 
$m(\eta_0)$ as the orbifolding group $\mathbb{Z}_N$ varies. For 
$N=3$, one sees that certain multiplicities vanish and thus there 
are fewer terms in the Fourier expansion for 
$\Delta_3(\mathbf{Z})$ as can be seen in appendix D.

The additive lift for $N=5$ is generated by a \textit{weak} 
Jacobi form. While the formula appears to go through, there are 
issues with the convergence of the series. This affects the 
holomorphicity of both the forms -- $\Phi_2(\mathbf{Z})$ and 
$\Delta_1(\mathbf{Z})$. However, these forms (and their modular 
transforms) appear to be compatible with the entropy of black 
holes as well as the higher-derivative $R^2$-corrections in the 
low-energy effective action. It is not clear to us whether this 
meromorphicity of $\Delta_1(\mathbf{Z})$ affects the existence of 
the GKM superalgebra $\mathcal{G}_5$ but we caution the reader 
about this possibility.

One may wonder what goes wrong when $N=7$. The weak Jacobi form has 
half-integral weight as well as half-integral index. It does not 
have an integral Fourier expansion and hence does not seem to 
lead to a GKM superalgebra with integer multiplicities. There is 
a related issue -- the modular form $\Phi_1(\mathbf{Z})$ is one 
with character and hence we anticipates subtleties associated 
with it.

\subsection{Interpreting the Jacobi form $\psi_{k/2,1/2}(z_1,z_2)$}

Given a Siegel modular form, $F_k({\mathbf Z})$ of weight $k$ 
with ${\mathbf Z}\in {\mathbb H}_2$, its Fourier expansion with 
respect to $z_3$ can be written as
\begin{equation}
F_k({\mathbf Z}) = \sum_{m=0}^\infty \phi_{k,m}(z_1,z_2)\ \exp(2\pi i m z_3)\ .
\end{equation}
For cusp forms, the term $m=0$ vanishes. The Fourier coefficients 
$\phi_{k,m}(z_1,z_2)$ are weak Jacobi forms of weight $k$ and 
index $m$ under the sub-group, $\Gamma^J$ of $Sp(2,\BZ)$ under 
which the cusp $\textrm{Im}(z_3)=\infty$ is invariant(see 
appendix \ref{JacobiForms}). For Siegel modular forms with 
character such as $\Delta_5({\mathbf Z})$, the above sum runs over 
half-integers and we obtain Jacobi forms with half-integral 
index.

In our situation, taking the limit 
$\textrm{Im}(z_3)\rightarrow\infty$ can also be understood as the 
removal of a real root of the GKM superalgebra ${\mathcal G}_N$. 
The Cartan matrix for the two-dimensional subspace of real roots 
is then the same as the affine Kac-Moody algebra 
$\hat{A}^{(1)}_1$ at level one. We thus see that 
$\hat{A}^{(1)}_1\subset {\mathcal G}_N$. As is well known and 
discussed earlier, the denominator formula for this algebra is 
given by $i \theta_1(z_1,z_2)$. In the product form for the theta 
function, it is easy to see the appearance of imaginary roots 
with multiplicity one. The Jacobi forms of $\Gamma_0(N)$ that we 
are interested in contain other powers of $\eta$ -- these are 
reflected in the fact that the $\mathcal{G}_N$ has more 
light-like imaginary roots than $\hat{A}^{(1)}_1$ -- it is these 
roots that appear in the correction terms in the sum side of the 
denominator formula. For instance, the imaginary simple root 
$\eta_0=2f_{-2}$ has multiplicity $m(\eta_0)=9$ for $N=1$. The 
same root appears in $L_+$ with multiplicity $10$. The difference 
is easy to understand -- the affine Lie algebra has one 
light-like root that is \textit{not} simple -- it is the sum of 
the two real roots. For $N=3,5$, we find $m(\eta_0)=0,-1$ respectively.
So we see that the root $\eta_0$ is a fermionic one for $N=5$.

As mentioned earlier, the analysis of David and Sen using the 
4D-5D lift leads to a separation of the product formula for 
$\Phi_k(\mathbf Z)$ into three terms. Taking the square-root to 
obtain $\Delta_{k/2}(\mathbf{Z})$ does not change the separation. 
We see that two of the terms are nothing but the product 
representation for the weak Jacobi form of index $\tfrac12$, 
$\psi_{k/2,1/2}(z_1,z_2)$. It appears that real and light-like 
simple roots appear from the spacetime and $T^2$ sectors in the 
type II picture. In particular, the electrically charged 
heterotic string states appear to arise from such light-like 
simple roots. We believe that this is a small step in 
understanding the connection between the GKM superalgebras and the 
algebra of BPS states.

\section{Conclusion and Outlook}

In this paper, we have shown that the square-root of the 
automorphic form $\Phi_k(\mathbf{Z})$ that generates the 
degeneracy of $1/4$-BPS CHL dyons can be interpreted as the 
Weyl-Kac-Borcherds denominator formula for a GKM superalgebra. 
Further, we have proposed an additive lift that directly 
generates the automorphic form $\Delta_k(\mathbf{Z})$ from a weak 
Jacobi form of index $1/2$. From the physical point of view, 
using the 4D-5D lift, we have been able to show that the real 
roots and light-like imaginary roots for electrically charged 
states arise from spacetime and $T^2$ sectors (in the type II 
picture) while the other imaginary roots necessarily arise from 
the $K3$ sector.

Cheng and Verlinde\cite{Cheng:2008fc} observe that the walls of 
the Weyl chambers for the GKM superalgebra $\mathcal{G}_1$ get 
mapped to walls of marginal stability for the $1/4$-BPS 
dyons\cite{Sen:2007vb,Sen:2007nz,Mukherjee:2007nc,Mukherjee:2007af}. 
This observation, if extended to the $\BZ_N$-orbifolds, seems, at 
first sight, to be in contradiction with our observation that the 
Weyl group remained unchanged for $\mathcal{G}_N$. However this 
naive extension is not quite correct, since for $N>1$, there are 
two distinct modular forms that have been constructed by 
Jatkar-Sen\cite{Jatkar:2005bh}, $\Phi_k(\mathbf{Z})$ and 
$\widetilde{\Phi}_k(\mathbf{Z})$. The first one is related to the 
$R^2$-corrections in the low-energy effective action while the 
second one is the one that generates the degeneracy of $1/4$-BPS 
dyons. Walls of marginal stability are precisely where this 
degeneracy jumps. The extension of the Cheng and Verlinde 
observation should be applicable to GKM superalgebras related to 
the square-root of $\widetilde{\Phi}_k(\mathbf{Z})$. This 
analysis has been recently carried out by Cheng and 
Dabholkar\cite{Cheng:2008kt} who find that the Weyl groups of the 
correspoding GKM superalgebras change with $N$ and the walls of 
the Weyl chambers get mapped to the walls of marginal stability 
for $N=2,3$.

Sen has studied the fundamental domain in the upper half-plane 
bounded by walls of marginal stability\cite{Sen:2007vb} and from 
his results one sees that for $N>3$, the domain has infinite 
volume (in the Poincar\'e metric). Cheng and Dabholkar also 
observe that the $N=5$ situation violates a certain finiteness 
condition (imposed by Gritsenko and Nikulin in their 
classification of rank three Lorentzian Kac-Moody 
algebras\cite{Nikulin:1996}) and thus they conclude that there is 
\textit{no} GKM superalgbra for $N=5$. Should one relax the 
finiteness condition and look for a GKM superalgebra for $N=5$?  
Is the meromorphicity of the modular form that we observed for 
$N=5$ relevant? We do not have any concrete answers to these 
questions and we leave if for future considerations.

Garbagnati and Sarti have studied symplectic (Nikulin) 
involutions of K3 
manifolds\cite{Garbagnati:2006,Garbagnati:2008}. In particular, 
they have explicitly constructed elliptic K3's whose automorphism 
groups are the Nikulin involutions. Further, they have provided 
an explicit description of the invariant lattice and its 
complementary lattice. We anticipate that these results might be 
relevant in improving our physical understanding the role of the 
roots of the GKM superalgebras.  The Jatkar-Sen construction 
holds for $N=11$ as well and it leads to a modular function 
(i.e., one of weight $k=0$) $\Phi_0(\mathbf{Z})$ and it is 
believed that a CHL string may exist. In the type IIA picture, 
the $\BZ_{11}$ is no longer a symplectic Nikulin involution, it 
acts non-trivially on $H^*(K3)$ and not on $H^{1,1}(K3)$ alone. 
It is of interest to study aspects of the $\BZ_N$ orbifold both 
from the physical and mathematical point of view.

As mentioned in the introduction, our aim has been to address the 
algebra of BPS states. While we have uncovered a nice algebraic 
structure, no direct relationship to the algebra of BPS states 
has been achieved. A related problem is that the degeneracy of 
BPS states appears related to the direct sum of two copies of the 
GKM superalgebras, $\mathcal{G}_N$. BPS states thus seem to be 
elements of a module that is a tensor product of two copies of 
(irreducible?) representations of the superalgebra. It would be 
nice to have a microscopic understanding of these issues. A 
possibly relevant observation here is that the elliptic genus of 
the Enriques surface\cite{Gritsenko:1999}, these are 
two-dimensional complex surfaces that arise as fixed-point free 
$\BZ_2$ involutions of K3 surfaces, gives rise to 
$\Delta_5({\mathbf Z})$. This might provide a hint on the 
appearance of two identical copies of the same GKM superalgebra 
for K3.

As we have seen, for affine Kac-Moody algebras, the presence of 
light-like imaginary roots in $L_+$ leads to powers of the 
Dedekind eta function appearing in the product form of the 
Weyl-Kac denominator formula. As is well known, 
$q^{1/24}/\eta(\tau)$ is the generating function of partitions of 
$n$ (equivalently, Young diagrams with $n$ boxes). An interesting 
generalisation is the generating function of plane partitions (or 
3D Young diagrams) has a nice product representation 
$\eta_{3D}\sim \prod_n (1-q^n)^n$ (due to MacMahon). This 
function appears in the counting of D0-branes in the work of 
Gopakumar-Vafa\cite{Gopakumar:1998ii,Gopakumar:1998jq}. Is there 
an algebraic interpretation for this? The addition of D2-branes 
to this enriches this story and leads to interesting 
formulae\cite{Young:2008}.

\bigskip \bigskip

\noindent \textbf{Acknowledgments:} We would like to thank Dileep 
Jatkar and Hiroki Aoki for patiently answering our numerous 
queries via email. KGK would like to thank S. Kalyana Rama for 
his support. SG would like to thank the CERN Theory Group for a 
visit during which some of this work was carried out. We thank 
the anonymous referee for critical comments that have led to 
several improvements in our paper.

\appendix
\section{Theta functions}
\subsection{Genus-one theta functions}\label{genusone}

The genus-one theta functions are defined by
\begin{equation}
\theta\left[\genfrac{}{}{0pt}{}{a}{b}\right] \left(z_1,z_2\right)
=\sum_{l \in \BZ} 
q^{\frac12 (l+\frac{a}2)^2}\ 
r^{(l+\frac{a}2)}\ e^{i\pi lb}\ ,
\end{equation}
where $a.b\in (0,1)\mod 2$. One has $\vartheta_1 
\left(z_1,z_2\right)\equiv\theta\left[\genfrac{}{}{0pt}{}{1}{1}\right] 
\left(z_1,z_2\right)$, $\vartheta_2 
\left(z_1,z_2\right)\equiv\theta\left[\genfrac{}{}{0pt}{}{1}{0}\right] 
\left(z_1,z_2\right)$, $\vartheta_3 
\left(z_1,z_2\right)\equiv\theta\left[\genfrac{}{}{0pt}{}{0}{0}\right] 
\left(z_1,z_2\right)$ and $\vartheta_4 
\left(z_1,z_2\right)\equiv\theta\left[\genfrac{}{}{0pt}{}{0}{1}\right] 
\left(z_1,z_2\right)$.

The transformations of $\vartheta_1(\tau,z)$ under modular transformations
is given by
\begin{eqnarray}
T:\qquad \quad\! \vartheta_1(\tau+1,z) &=& e^{i\pi/4}\
\vartheta_1(\tau,z)\ ,
\nonumber \\
S:\quad  \vartheta_1(-1/\tau,-z/\tau) &=& -\frac{1}{q^{1/2}r}\ e^{\pi
iz^2/\tau}\ \vartheta_1(\tau,z)\ ,
\end{eqnarray}
with $q=\exp(2\pi i \tau)$ and $r=\exp(2\pi i z)$.

\noindent
The Dedekind eta function $\eta(\tau)$ is defined by
\begin{equation}
%\label{ }
\eta(\tau)= e^{2\pi i \tau/24} \prod_{n=1}^\infty (1-q^n)\ .
\end{equation}
The transformation of the Dedekind eta function under the modular 
group is given by
\begin{eqnarray}
T:\quad \eta(\tau+1) & = & e^{\pi i/12}\ \eta(\tau)\ , \nonumber \\
S:\quad \eta(-1/\tau) & = & e^{-\pi i/4}\ (\tau)^{1/2}\ \eta(\tau) \ .
\end{eqnarray}
The transformation of $\eta(N\tau)$ is given by
\begin{eqnarray}
T:\quad \eta(N\tau+N) & = & e^{N \pi i/12}\ \eta(\tau)\ , \nonumber \\
S:\qquad \ \eta(-1/\tau) & = & \frac{e^{-\pi i/4}}{\sqrt{N}}\ (\tau)^{1/2}\ 
\eta(\tau/N) \ .
\end{eqnarray}
One can see that $\eta(N\tau)$ transforms into $\eta(\tau/N)$ 
under the $S$ transformation. $\eta(N\tau)$ gets mapped to itself 
only under the subgroup, $\Gamma_0(N)$ of $SL(2,\BZ)$. Following 
Niemann\cite{Niemann}, let
\begin{equation}
%\label{ }
\psi_j(\tau)\equiv \eta\left(\tfrac{\tau+j}{N} +j \right)\ ,\ 
j=0,1,\ldots, N-1\mod N\ .
\end{equation}
Both $S$ and $T$ no longer have a diagonal action on the 
$\psi_j(\tau)$. One has
\begin{eqnarray}
T:\quad \psi_j(\tau+1) & = & e^{\pi i/12}\ \psi_{j+1}(\tau)  \\
S:\quad \psi_j(-1/\tau) & = & e^{(j+j')\pi i/12}\ (\tau)^{1/2}\ 
\chi(G)\ \psi_{-j'}(\tau)\ ,
\end{eqnarray}
where $jj'=1\mod N$ and the character $\chi(G)$ has to be 
calculated on a case by case basis (see chapter 2 of 
\cite{Niemann} for details).

The transformations of the eta related functions show us that the 
functions $f^{k}(\tau)$ and its square root can transform with 
non-trivial character. In particular, one can show that for 
$N=7$, $f^{(1)}(\tau)$ and for $N=3$, $\sqrt{f^{(4)}(\tau)}$ 
transform with character. As these two functions enter the weak 
Jacobi forms that are used to construct the Siegel modular forms 
$\Phi_1(\mathbf{Z})$ and $\Delta_2(\mathbf{Z})$ respectively, 
these two Siegel modular forms will transform with non-trivial 
character\cite{Eichler}. This is the basis for our claim that 
$\Delta_2(\mathbf{Z})$ must transform with non-trivial character 
and is consistent with the observation of Jatkar-Sen regarding 
$\Phi_1(\mathbf{Z})$.

\subsection{Genus-two theta constants}

We define the genus-two theta constants as follows\cite{Nikulin:1995}:
\begin{equation}
\theta\left[\genfrac{}{}{0pt}{}{\mathbf{a}}{\mathbf{b}}\right]
\left(\mathbf{Z}\right)
=\sum_{(l_1, l_2)\in \BZ^2} 
q^{\frac12 (l_1+\frac{a_1}2)^2}\ 
r^{(l_1+\frac{a_1}2)(l_2+\frac{a_2}2)}\ 
s^{\frac12 (l_2+\frac{a_2}2)^2}\ 
e^{i\pi(l_1b_1+l_2b_2)}\ ,
\end{equation}
where $\mathbf{a}=\begin{pmatrix}a_1\\ a_2
\end{pmatrix}$,
$\mathbf{b}=\begin{pmatrix}b_1\\ b_2
\end{pmatrix}$,
and $\mathbf{Z}=\begin{pmatrix}z_1 & z_2 \\ z_2 &
z_3\end{pmatrix}\in \mathbb{H}_2$. 
Further, we have defined $q=\exp(2\pi i z_1)$,
$r=\exp(2\pi i z_2)$ and $s=\exp(2\pi i z_3)$.
The constants $(a_1,a_2,b_1,b_2)$ take values $(0,1)$. 
Thus there are sixteen genus-two theta constants. The
even theta constants are those for which
$\mathbf{a}^{\textrm{T}}\mathbf{b}=0\mod 2$. There are ten such theta
constants for which we list the values of $\mathbf{a}$ and $\mathbf{b}$:
\begin{center}
\begin{tabular}{|c|c|c|c|c|c|c|c|c|c|c|}\hline 
$m$ & 0 & 1 & 2 & 3 & 4 & 5 & 6 & 7 & 8 & 9 
\\ \hline 
$\begin{pmatrix} \mathbf{a} \\ \mathbf{b} \end{pmatrix}$ &
$\left(\begin{smallmatrix}0\\0\\0\\0 \end{smallmatrix}\right)$ &
$\left(\begin{smallmatrix}0\\1\\0\\0 \end{smallmatrix}\right)$ &
$\left(\begin{smallmatrix}1\\0\\0\\0 \end{smallmatrix}\right)$ &
$\left(\begin{smallmatrix}1\\1\\0\\0 \end{smallmatrix}\right)$ &
$\left(\begin{smallmatrix}0\\0\\0\\1 \end{smallmatrix}\right)$ &
$\left(\begin{smallmatrix}1\\0\\0\\1 \end{smallmatrix}\right)$ &
$\left(\begin{smallmatrix}0\\0\\1\\0 \end{smallmatrix}\right)$ &
$\left(\begin{smallmatrix}0\\1\\1\\0 \end{smallmatrix}\right)$ &
$\left(\begin{smallmatrix}0\\0\\1\\1 \end{smallmatrix}\right)$ &
$\left(\begin{smallmatrix}1\\1\\1\\1 \end{smallmatrix}\right)$  \\ \hline
\end{tabular}
\end{center}
We will refer to the above ten theta constants as $\theta_m(\mathbf{Z})$
with $m=0,1,\ldots,9$ representing the ten values of 
$\mathbf{a}$ and $\mathbf{b}$ as defined in the above table.

The modular functions $\Delta_5(\mathbf{Z})$ and 
$\Delta_3(\mathbf{Z})$ can be written out in terms of the even 
theta constants\cite{Nikulin:1995,Raghavan}. One finds
\begin{eqnarray}
\Delta_5(\mathbf{Z}) &=& \frac1{64} \prod_{m=0}^9 \theta_m(\mathbf{Z})\ , \\
\Delta_3(\mathbf{Z}) &=& \frac1{64}\ \theta_2(\mathbf{Z})\! 
\prod_{m=1\textrm{ mod }2} \theta_m(\mathbf{Z})\ .
\end{eqnarray}
Following ref. \cite{Raghavan}, the formula for $\Delta_3(\mathbf{Z})$ was 
constructed experimentally by looking for a product of six even 
theta constants that had the correct series expansion. This is 
also in agreement with the expression for $\Phi_6(\mathbf{Z})$ given in ref. 
\cite{Aoki:2005} in terms of theta constants.

Let us define $\tilde{\Delta}_3(\mathbf{Z})$ to be the square-root of 
$\tilde{\Phi}_6(\mathbf{Z})$. One may wonder if it can also be written as the 
product of six genus-two theta constants. Keeping in mind that 
the leading term will go as $q^{1/4}r^{1/2}s^{1/2}$, we find that 
the following combination achieves this (to the orders that we 
have verified):
\begin{equation}
\tilde{\Delta}_3(\mathbf{Z}) = \frac1{16}\ \theta_1(\mathbf{Z}) \
\theta_3(\mathbf{Z})\ 
\theta_6(\mathbf{Z})\ 
\theta_7(\mathbf{Z})\ 
\theta_8(\mathbf{Z})\ 
\theta_9(\mathbf{Z})\ ,
\end{equation}
squares to given $\tilde{\Phi}_6(\mathbf{Z})$.

\section{Jacobi cusp forms}\label{JacobiForms}

\subsection{Basic group theory}

The group $Sp(2,\BZ)$ is the set of $4\times 4$ matrices written 
in terms of four $2\times 2$ matrices $A,\, B,\, C,\, D$ 
as\footnote{This section is based on the book by Eichler and 
Zagier\cite{Eichler}.}
$$
M=\begin{pmatrix}
   A   & B   \\
    C  &  D
\end{pmatrix}
$$
satisfying $ A B^T = B A^T $, $ CD^T=D C^T $ and $ AD^T-BC^T=I $. 
The congruence subgroup $\widehat{G}_0(N)$ of $Sp(2,\BZ)$ is given by the 
set of matrices such that $C=0\mod N$. This group acts naturally 
on the Siegel upper half space, $\BH_2$, as
\begin{equation}
%\label{ }
\mathbf{Z}=\begin{pmatrix} z_1 & z_2 \\ z_2 & z_3 \end{pmatrix}
\longmapsto M\cdot \mathbf{Z}\equiv (A \mathbf{Z} + B) 
(C\mathbf{Z} + D)^{-1} \ .
\end{equation}

The Jacobi group $\Gamma^J =SL(2,\BZ)\ltimes H(\BZ)$ is the
sub-group of $Sp(2,\BZ)$ that preserves the one-dimensional cusp
$z_3=i \infty$. The $SL(2,\BZ)$ 
is generated by the embedding of 
$\left(\begin{smallmatrix} a & b \\ c & d\end{smallmatrix}\right)
\in SL(2,\BZ)$ in $Sp(2,\BZ)$ 
\begin{equation}
\label{sl2embed}
g_1(a,b;c,d)\equiv \begin{pmatrix}
   a   &  0 & b & 0   \\
     0 & 1 & 0 & 0 \\
     c &  0 & d & 0 \\
     0 & 0 & 0 & 1  
\end{pmatrix}
\ .
\end{equation}
The above matrix acts on $\BH_2$ as
\begin{equation}
(z_1,z_2,z_3) \longrightarrow \left(\frac{a z_1 + b}{cz_1+d},\  
\frac{z_2}{cz_1+d},\  z_3-\frac{c z_2^2}{c z_1+d}\right)\ ,
\end{equation}
with $\det(C\mathbf{Z} + D)=(cz_1+d)$. The Heisenberg group, 
$H(\BZ)$, is generated by $Sp(2,\BZ)$ matrices of the form
\begin{equation}
\label{sl2embedapp}
g_2(\lambda, \mu,\kappa)\equiv \begin{pmatrix}
   1   &  0 & 0 & \mu   \\
    \lambda & 1 & \mu & \kappa \\
     0 &  0 & 1 & -\lambda \\
     0 & 0 & 0 & 1  
\end{pmatrix}
\qquad \textrm{with } \lambda, \mu, \kappa \in \BZ
\end{equation}
The above matrix acts on $\BH_2$ as
\begin{equation}
(z_1,z_2,z_3) \longrightarrow \left(z_1,\ \lambda z_1 + z_2 + \mu,\  
z_3+ \lambda^2 z_1 + 2 \lambda z_2 + \lambda \mu \right)\ ,
\end{equation}
with $\det(C\mathbf{Z} + D)=1$. It is easy to see that $\Gamma^J$ 
preserves the one-dimensional cusp at $\textrm{Im}(z_3)= \infty$.

The full group $Sp(2,\BZ)$ is generated by adding the exchange element
to the group $\Gamma^J$.
\begin{equation}
g_3 \equiv 
\begin{pmatrix}
   0   &  1 & 0 & 0   \\
     1 & 0 & 0 & 0 \\
     0 &  0 & 0 & 1 \\
     0 & 0 & 1 & 0 
\end{pmatrix}
\ .
\end{equation}
This acts on $\BH_2$ exchanging $z_1\leftrightarrow z_3$. The 
subgroup $\widehat{G}_0(N)$ is generated by considering the same three sets 
of matrices with the additional condition that 
$\left(\begin{smallmatrix} a & b \\ c & 
d\end{smallmatrix}\right)\in \Gamma_0(N)$ i.e., $c=0\mod N$ in 
Eq. (\ref{sl2embed}). Further, we will call the corresponding Jacobi group $\Gamma_0(N)^J$.

\subsection{Weak Jacobi forms}

A \textit{Siegel modular form} of weight $k$ and character $v$ 
with respect to $Sp(2,\BZ)$ is a holomorphic function $F: \BH_2 
\rightarrow \BC$ satisfying
\begin{equation}
F(M\cdot \mathbf{Z}) = v(M)\ \det(C\mathbf{Z}+D)^k \ F(\mathbf{Z})\ ,
\end{equation} 
for all $\mathbb{Z}\in \BH_2$ and $M\in Sp(2,\BZ)$. In the above definition, 
one can replace $Sp(2,\BZ)$ by any of its sub-groups such as $\Gamma^J$ or 
$\widehat{G}_0(N)$.

\noindent A holomorphic function 
\[
\phi_{k,t}(z_1,z_2):\quad \BH_1\times \BC \rightarrow \BC\ ,
\]
 is 
called a \textit{Jacobi form} of weight $k$ and index $t\in \frac12\BZ$, if
\begin{enumerate}
\item The function 
$$
\tilde{\phi}_k(\mathbf{Z})=\exp(2\pi i t z_3)\ \phi_{k,t}(z_1,z_2)\ ,
$$ 
on $\BH_2$ is a modular form of
weight $k$ with respect to the Jacobi group $\Gamma^J \subset
Sp(2,\BZ)$.
\item It has a Fourier expansion
\begin{equation}
\phi_{k,t}(z_1,z_2)= \sum_{\ell\in t+\BZ } f(n,\ell)\
q^n r^\ell \ ,
\end{equation}
such that the Fourier coefficients, $f(n,\ell)=0$, unless $n\geq0$ and
$4nt-\ell^2\geq0$.
\end{enumerate}
For \textit{weak} Jacobi forms, the coefficients $f(n,\ell)$ are 
non-vanishing only when $n\geq0$ relaxing the condition involving 
$(4nt-\ell^2)$. Jacobi forms of integer index were considered by 
Eichler and Zagier\cite{Eichler} and extended to half-integral 
indices by Gritsenko\cite{Gritsenko:1999}.

The elliptic genus of Calabi-Yau manifolds are weak Jacobi forms. 
Examples include:
\begin{eqnarray}
\phi_{-2,1}(z_1,z_2)&=&\mathcal{E}_{\textrm{st}\times T^2}(z_1,z_2)
=\left(\frac{i\vartheta_1(z_1,z_2)}{\eta^3(z_1)}\right)^2 \nonumber \\
\phi_{0,1}(z_1,z_2)&=&\mathcal{E}_{K3}(z_1,z_2) 
= 8 \sum_{i=2}^4 \left(\frac{\vartheta_i(z_1,z_2)}{\vartheta_i(z_1,0)}\right)^2 
\label{weakJacobiForm}
\end{eqnarray}

We will see the appearance of weight zero Jacobi forms of the 
group $\Gamma_0(N)^J$ in writing product representations for the 
modular form $\Phi_k(\mathbf{Z})$.
\begin{equation}
\phi^{(N)}_{0,1}(\tau,z)= \frac{2N}{N+1} \ \alpha^{(N)}(\tau)\ 
\phi_{-2,1}(\tau,z) + \frac1{N+1}\ \phi_{0,1}(\tau,z)\ ,
\end{equation}
with $\alpha^{(N)}(\tau)=\tfrac{12i}{\pi(N-1)}\partial_\tau 
\big[\ln \eta(\tau) -\ln \eta(N\tau)\big]$ is the Eisenstein 
series for $\Gamma_0(N)$. The Fourier expansion for 
$\phi^{(N)}_{0,1}$ at the cusp at $i\infty$
\begin{align}
  \phi^{(2)}_{0,1}(\tau,z)= &\left(2 r+4+\frac{2}{r}\right)+\left(4
   r^2-8+\frac{4}{r^2}\right)q + O\left(q^{2}\right)    \nonumber \\
 \phi^{(3)}_{0,1}(\tau,z)   =& \left(2 r+2+\frac{2}{r}\right)+\left(2 r^2-2
   r-\frac{2}{r}+\frac{2}{r^2}\right) q+O\left(q^{2}\right) \\
 \phi^{(5)}_{0,1}(\tau,z) =& \left(2 r+\frac{2}{r}\right)+\left(2
   r-4+\frac{2}{r}\right) q+O\left(q^{2}\right) .\nonumber
\end{align}
and about the cusp at $0$ is
\begin{align}
%\label{}
  \phi^{(2)}_{0,1}= & 8+\left(-\frac{16}{r}+32-16
   r\right)
   q^{1/2}
  +\left(\frac{8}{r^2}-\frac{64}{r}+112-64 r+8 r^2\right)q
   + O\left(q^{3/2}\right) \nonumber   \\
 \phi^{(3)}_{0,1}=&6+\left(-\frac{6}{r}+12-6
   r\right)
   q^{1/3}+\left(-\frac{18}{r}+36-18
   r\right)
   q^{2/3}\nonumber \\ 
   &+\left(\frac{6}{r^2}-\frac{42}{r}+72-42 r+6
   r^2\right)
   q+O\left(q^{4/3}\right) \\
 \phi^{(5)}_{0,1}=&4+\left(-\frac{2}{r}+4-2
   r\right)
   q^{1/3}+\left(-\frac{6}{r}+12-6
   r\right)
   q^{2/5}+\left(-\frac{8}{r}+16-8
   r\right)
   q^{3/5} \nonumber \\
   &+\left(-\frac{14}{r}+28-14
   r\right)
   q^{4/5}+\left(\frac{4}{r^2}-\frac{26}{r}+44-26 r+4
   r^2\right)
   q +O\left(q^{6/5}\right) . \nonumber
\end{align}

\subsection{Additive lift of Jacobi forms with integer index}

Given a Jacobi form of weight $k$ and index $1$, Maa\ss\ constructed a
Siegel modular form of weight $k$ leading to an explicit formula\cite{Maass}
\begin{equation}
\Phi_{k}(\mathbf{Z}) \equiv \sum_{(n,\ell,m)>0}\ \  
\sum_{d | (n,\ell,m)} \ 
d^{k-1}\ f\left(\tfrac{nm}{d^2},\tfrac\ell{d}\right)\ 
q^{n} r^{\ell} s^{m}\ .
\end{equation}

This procedure is known as the \textit{arithmetic} or 
\textit{additive} lift of the Jacobi form. It is known that the 
ring of Siegel modular forms is generated by four modular forms 
with weights $4$, $6$, $10$ and $12$. For instance, the weight $10$ 
modular form, $\Phi_{10}(\mathbf{Z})$, is generated by the Jacobi 
form of weight $10$ and index $1$
\begin{equation}
\phi_{10,1}(z_1,z_2) = \theta_1(z_1,z_2)^2\  \eta(z_1)^{18}\ .
\end{equation}

Given a weight $k$, index $1$ Jacobi form of the subgroup 
$\Gamma_0(N)^J$, one has a similar formula leading to a level-$N$ 
Siegel modular form i.e, a modular form of $\widehat{G}_0(N)$, is 
given by\cite{Manickam:1993,Manickam:2002,Aoki:2008}
\begin{equation}\label{Maasswithcharacter}
\Phi_{k}(\mathbf{Z}) \equiv \sum_{(n,\ell,m)>0}\ \  
\sum_{d | (n,\ell,m)} \ \chi(d)\ 
d^{k-1}\ f\left(\tfrac{nm}{d^2},\tfrac\ell{d}\right)\ 
q^{n} r^{\ell} s^{m}\ .
\end{equation}
where $\chi(d)$ is a real Dirichlet character\cite{Dirichletchar} 
modulo $N$. When $\chi(d)$ is trivial, i.e.,
\begin{equation}\label{trivialchar}
\chi(d)=\left\{\begin{array}{ll}0&  \textrm{if } (d,N)\neq 1 \\ 1 & 
\textrm{otherwise }, \end{array}\right.
\end{equation}
we obtain a level $N$ Siegel modular form. For levels $N=2,3,5$, 
Jatkar and Sen have constructed modular forms of weight $k=6,4,2$ 
respectively as additive lifts of the weight $k$ Jacobi 
forms\cite{Jatkar:2005bh}:
\begin{equation}
\phi_{k,1}(z_1,z_2)=\theta_1(z_1,z_2)^2 \ \eta(z_1)^{k-4}\ \eta(Nz_1)^{k+2}\ .
\end{equation}

When the Jacobi form is one with character, one sees the 
appearance of a non-trivial Dirichlet character and the Siegel 
modular form obtained from the additive lift is one with 
character. At level $N=7$, Jatkar and Sen have constructed a 
modular form of weight $1$ with character\cite{Jatkar:2005bh}. In 
this case, one does indeed see the appearance of a non-trivial 
real Dirichlet character in Eq. (\ref{Maasswithcharacter}) given 
above.

\section{The additive lift with character at level $N$}\label{modularity}

Let $\Delta_k(\mathbf{Z})$ be a Siegel modular form of 
$\widehat{G}_0(N)\subset Sp(2,\BZ)$ with character $v^\Gamma$ 
i.e.,
\begin{equation}
\Delta_k(M\cdot \mathbf{Z})= v^\Gamma(M)\ \det(C\mathbf{Z}+D)^k\  \Delta_k(\mathbf{Z})\ ,
\end{equation}
where $v^\Gamma(M)$ is the unique non-trivial real linear 
character of $Sp(2,\BZ)$\cite{Reiner:1955} and 
$M=\left(\begin{smallmatrix}A & B \\ C & 
D\end{smallmatrix}\right)\in \widehat{G}_0(N)$. An explicit 
expression for $v^\Gamma(M)$ is\cite{Maass:1980}
\begin{align}
 v^\Gamma\begin{pmatrix}0 & -I_2 \\ I_2 & 0 \end{pmatrix} =1 \ ,\ &
v^\Gamma\begin{pmatrix}I_2 & B \\ 0 & I_2 \end{pmatrix} =(-1)^{b_1+b_2+b}\ , \\
v^\Gamma\begin{pmatrix}U^T & 0 \\ 0 & U^{_1} \end{pmatrix}
&=(-1)^{(1+u_0+u_2)(1+u_1+u_3)+u_0u_2}\ ,
\end{align}  
where $I_2$ is the $2\times 2$ identity matrix, 
$B=\left(\begin{smallmatrix}b_1 & b \\ b & 
b_2\end{smallmatrix}\right)$ and $U=\left(\begin{smallmatrix}u_0 
& u_3 \\ u_1 & u_2\end{smallmatrix}\right)$ is a uni-modular 
matrix. The embedding of $SL(2,\BZ)$ in $Sp(2\BZ)$ given in Eq. 
\eqref{sl2embed} induces a character for $SL(2,\BZ)$. Given a 
$SL(2,\BZ)$ matrix $M$, let $M^*$ denote the corresponding 
$Sp(2,\BZ)$ matrix as determined by Eq. \eqref{sl2embed}, one has
\begin{equation}
 w^\Gamma(M)\equiv v^\Gamma(M^*)\ ,
\end{equation} 
where $w^\Gamma(M)$ is the $SL(2,\BZ)$ character induced by the 
character of $Sp(2,\BZ)$. It is useful to note that 
$w^\Gamma(S)=1$ when $S=I_2\textrm{ mod } 2$.

Such a modular form admits a Fourier-Jacobi expansion of the form
\begin{equation}
\Delta_k(\mathbf{Z})= \sum_{m=1\textrm{ mod }2; m>0} \psi_{k,m/2}(z_1,z_2)\ 
s^{m/2}\ ,
\end{equation}
where $\psi_{k,m/2}(z_1,z_2)$ is a Jacobi form of weight $k$ and 
index $m/2$ and character $w^\Gamma$ (induced by $v^\Gamma)$. The 
condition that the exponent in the Fourier expansion be 
half-integral follows from the behavior of the character. 
Similarly, one also has the following transformations of 
$\psi_{k,m/2}(z_1,z_2)$
\begin{align}
\psi_{k,m/2}(z_1,z_2+h) &= (-1)^h\ \psi_{k,m/2}(z_1,z_2)\ ,\\
\psi_{k,m/2}(z_1,\lambda z_1+ z_2) &= (-1)^\lambda\ \psi_{k,m/2}(z_1,z_2) \ .
\end{align}
Thus, one sees that Fourier expansion of the Jacobi form about 
the cusp at $i\infty$ has half-integral exponents. An important 
point to note is that the weight $k$ \textit{must} be odd else 
the modular form vanishes\cite{Maass:1980}.

\subsection{Additive lift of Jacobi forms with half-integer index}

So far we have only considered examples of modular forms obtained 
from the lift of Jacobi forms with integral index. We will now 
consider examples with half-integral index as they appear in the 
denominator for the the GKM algebras, $\mathcal{G}_N$ discussed 
in the main body of the paper. The simplest example is given by 
the Jacobi theta function, $\vartheta_1(z_1,z_2)$. It is a 
holomorphic Jacobi form of weight $1/2$ and index $1/2$ with 
character. This Jacobi form appears as the denominator formula of 
the affine Kac-Moody algebra, $\widehat{A}_1^{(1)}$.

The Jacobi form of weight $5$ and index $1/2$
\begin{equation}
 \psi_{5,1/2}(z_1,z_2) = \vartheta_1(z_1,z_2) \ \eta(z_1)^9\ ,
\end{equation}  
generates the Siegel modular form with character, 
$\Delta_5(\mathbf{Z})$ via the additive lift. The Fourier 
expansion of the Jacobi form now involves half-integral 
exponents. One has
\begin{equation}
 \psi_{5,1/2}(z_1,z_2) = \sum_{n,\ell=1\textrm{ mod }2} g(n,\ell)\ q^{n/2}
r^{\ell/2}\ ,
\end{equation}  
with $g(n,\ell)=0$ unless $4n-\ell^2\geq0$. The modular form
$ \Delta_5(\mathbf{Z})$ has the following expansion\cite{Maass:1980}
\begin{equation}
\label{Cten}
 \Delta_5(\mathbf{Z})= \sum_{(n,\ell,m)>0}\ \  
\sum_{d | (n,\ell,m)} \ 
d^{k-1}\ g\left(\tfrac{nm}{d^2},\tfrac\ell{d}\right)\ 
q^{n/2} r^{\ell/2} s^{m/2}\ .
\end{equation}
Notice the similarity with the Maa\ss\ formula given in Eq. 
(\ref{Maasswithcharacter}) with half-integral powers of $q$, $r$ 
and $s$ appearing where integral powers appeared. Gritsenko and 
Nikulin have shown that this modular form appears as the 
denominator formula of a GKM superalgebra.

$\Delta_5(\mathbf{Z})$ is a modular form with character under the 
full modular group, $Sp(2,\BZ)$. It transforms as
\begin{equation}
%\label{ }
\Delta_5(M\cdot \mathbf{Z}) = v^\Gamma(M) \ (C\mathbf{Z}+D)^5\  
\Delta_5(\mathbf{Z})\ ,
\end{equation}
We will now generalize the additive lift \eqref{Cten} to higher 
levels. In particular, we construct additive lifts of the Jacobi 
forms of index $1/2$ given in Eq. \eqref{AddJacobiForm}.

\subsection{Hecke operators and the additive lift}

Given a index-half Jacobi form $\psi_{k,1/2}(\tau,z)$ with Fourier expansion
\begin{equation}
\psi_{k,1/2}(\tau,z)= \sum_{n,\ell} \ g(n,\ell)\ q^{n/2} r^{\ell/2}\ ,
\end{equation}
we construct a Jacobi form of the same weight and index $m/2$ for 
odd $m$ by the following averaging procedure. This generalizes 
and closely follows the procedure due to Maa\ss\cite{Maass:1980} 
for Jacobi forms of index half to higher levels.

Let $X_m$ be the group of $2\times 2$ matrices $M=\left( 
\begin{smallmatrix}\alpha & \beta \\ \gamma & \delta 
\end{smallmatrix}\right)$ with integral entries such that (i) 
$\det(M)=m$ ($m=1\textrm{ mod }2$); (ii) $M=I_2 \mod 2$; (iii) 
$\gamma=0 \mod N$ and (iv) $(\alpha,N)=1$. Note that $X_1$ is a 
subgroup of $\Gamma_0(N)$ for which 
$w^\Gamma(M)=v^\Gamma(M^*)=1$. We will show that the following is 
a Jacobi form of weight $k$ and index $m/2$:
\begin{equation}
\psi_{k,m/2}(\tau,z) \equiv (m)^{k-1} \!\!\!\!\!
\sum_{\left(\begin{smallmatrix}\alpha & \beta \\ 
\gamma & \delta \end{smallmatrix}\right)\in X_1\backslash X_m}\!\!\! 
(\gamma \tau + \delta)^{-k} \ e^{\pi i m 
\frac{\gamma z^2}{\gamma \tau + \delta}}\ \psi_{k,1/2}
\left(\tfrac{\alpha\tau+\beta}{\gamma \tau+\delta}, 
\tfrac{mz}{\gamma\tau+\delta}\right)
\end{equation}

\noindent \textbf{Claim:} Consider the coset $X_1\backslash X_m$. 
The coset can be represented by the elements
\begin{equation}
 \begin{pmatrix}\alpha & 2 \beta \\ 0 & \delta \end{pmatrix}\ , 
\textrm{ with } \alpha \delta =m,\ \beta=0,1,\ldots, (\delta-1)\ , 
(\alpha, N)=1 \textrm{ and } \alpha>0\ .
\end{equation}
The proof is along lines similar to the one given by 
Jatkar-Sen\cite[see appendix A]{Jatkar:2005bh} and will not be 
repeated here. Using the above representation of the coset 
$X_1\backslash X_m$ given above, we can rewrite the equation 
defining $\psi_{k,m/2}(\tau,z)$ as follows:
\begin{align}
\psi_{k,m/2}(\tau,z) &= (m)^{k-1} 
\sum_{\tiny \begin{array}{c}\alpha\delta=m\\ 
(\alpha,N)=1\\ \alpha>0\end{array}} 
\delta^{-k}\ \sum_{\beta=0}^{\delta-1} 
\psi_{k,1/2}\left(\tfrac{\alpha\tau+2\beta}{\delta},\tfrac{mz}{\delta}\right)\\
&=(m)^{k-1} \sum_{\tiny \begin{array}{c}\alpha\delta=m\\ (\alpha,N)=1\\ 
\alpha>0\end{array}} \delta^{-k}\ \sum_{\beta=0}^{\delta-1} 
\sum_{n,\ell} g(n,\ell)\ e^{i\pi n \tfrac{\alpha\tau+2\beta}{\delta}} 
e^{i\pi\ell \tfrac{mz}{\delta} }\\
&=(m)^{k-1} \sum_{\tiny \begin{array}{c}\alpha\delta=m\\ (\alpha,N)=1\\ 
\alpha>0\end{array}} \delta^{-k+1}\ 
\sum_{\tiny \begin{array}{c}n,\ell \\ \delta | n\alpha\end{array}} g(n,\ell)\
e^{\tfrac{i\pi n \alpha\tau}{\delta}} e^{ \tfrac{i\pi \ell mz}{\delta} }\\
&=\sum_{n',\ell'}\ 
\sum_{\tiny \begin{array}{c}\alpha | (n',\ell',m)\\ \alpha>0 \end{array}}  
\chi(\alpha) \ 
\alpha^{k-1}\ g\left(\tfrac{n'm}{\alpha^2},\tfrac{\ell'}{\alpha}\right)\ 
\ e^{i\pi
n'\tau} e^{i\pi \ell' z }\ ,
\end{align}
where $\chi(\alpha)$ is the trivial Dirichlet character modulo 
$N$ which implements the condition $(\alpha,N)=1$. In obtaining 
the last line, we have defined $n' = n\alpha/\delta$ and 
$\ell'=\ell \alpha$ and replaced $\delta$ by $m/\alpha$.

\noindent\textbf{Modularity:} We now prove that 
$\psi_{k,m/2}(\tau,z)$ constructed by the averaging procedure is 
a Jacobi form with character. Let us first consider the 
transformation under $M_1 =\left(\begin{smallmatrix}a & b \\ c & 
d \end{smallmatrix}\right)\in \Gamma_0(N)$. The proof follows 
along the lines of Maa\ss\cite{Maass:1980} (and 
Jatkar-Sen\cite{Jatkar:2005bh} for the modifications to account 
for the level.). We only repeat the crucial argument. Given any 
matrix $S\in X_1\backslash X_m$, one has $S M_1 = \tilde{M}_1 
\tilde{S}$ where $\tilde{M}_1=\left(\begin{smallmatrix}\tilde{a} 
& \tilde{b} \\ \tilde{c} & \tilde{d} \end{smallmatrix}\right)\in 
\Gamma_0(N)$ and $\tilde{S}$ is another matrix in $X_1\backslash 
X_m$. In other words, averaging over all $S$ can be replaced by 
averaging over $\tilde{S}$ with $\tilde{M}_1$ taking the place of 
$M_1$. In particular, one has
\begin{equation}\label{Jacobichartransform}
\psi_{k,1/2}\left(\tfrac{\tilde{a} \tau + \tilde{b}}{\tilde{c}\tau+\tilde{d}}, 
\tfrac{z}{\tilde{c}\tau+\tilde{d}}\right) 
e^{-i\pi \tfrac{\tilde{c}z^2}{\tilde{c}\tau+\tilde{d}}} 
= w^\Gamma(\tilde{M}_1)\ (\tilde{c}\tau+\tilde{d})^k\  \psi_{k,1/2}(\tau,z)\ .
\end{equation}
Since $w^\Gamma(S M_1) = w^\Gamma(\tilde{M}_1 \tilde{S})$ and 
$w^\Gamma(S)=w^\Gamma(\tilde{S})=1$, we see that 
$w^\Gamma(M_1)=w^\Gamma(\tilde{M}_1)$ leading to the required 
result.
\begin{equation}
\psi_{k,m/2}\left(\tfrac{a \tau + b}{c\tau+d}, \tfrac{z}{c\tau+d}\right) e^{-i\pi m \tfrac{cz^2}{c\tau+c}} = w^\Gamma(M_1)\ (c\tau+d)^k\  \psi_{k,m/2}(\tau,z)\ .
\end{equation}
The transformation under the other generators of the Jacobi group,
$\Gamma_0(N)^J$, follow in an elementary fashion. Thus, we see that
$\psi_{k,m/2}$ is a Jacobi form of weight $k$ and index $m/2$ for odd $m$.

\noindent One thus sees that
\begin{align}\label{halfintegerlift}
\Delta_k(\mathbf{Z})&= \sum_{m=1\textrm{ mod }2; m>0}
\psi_{k,m/2}(z_1,z_2)\ e^{i\pi m z_3}\nonumber  \\
&=\sum_{(n,\ell,m)>0}\ \  
\sum_{\tiny\begin{array}{c}\alpha | (n,\ell,m)\\ \alpha>0 \end{array}
} \ \chi(\alpha)\ 
\alpha^{k-1}\ g\left(\tfrac{nm}{\alpha^2},\tfrac\ell{\alpha}\right)\ 
q^{n/2} r^{\ell/2} s^{m/2}\ .
\end{align}
is a modular form of $\widehat{G}_0(N)$ with (odd) weight $k$ and 
character $v^\Gamma$. When the seed Jacobi form is a weak one, 
then the above sum is \textit{not} convergent and one obtains a 
meromorphic modular form\cite{Aoki:2008}. This is indeed the case 
in our example for $N=5$.

When $N=3$, the weight of the modular form we have constructed is 
even as $k=2$. However, the seed Jacobi form, 
$\psi_{2,1/2}(\tau,z)$, transforms with character $w^\gamma 
w^\psi$ and not as in Eq. \eqref{Jacobichartransform}, thus 
evading the restriction on $k$ being odd. Taking into account the 
additional character, $w^\psi$, one obtains a formula similar to 
Eq. \eqref{halfintegerlift} but with $\chi^\psi(\alpha)$ as 
defined in Eq. \eqref{modthree} replacing $\chi(\alpha)$:
\begin{equation}
\Delta_2(\mathbf{Z})
=\sum_{(n,\ell,m)>0}\ \  
\sum_{\tiny\begin{array}{c}\alpha | (n,\ell,m)\\ \alpha>0 \end{array}
} \ \chi^\psi(\alpha)\ 
\alpha^{k-1}\ g\left(\tfrac{nm}{\alpha^2},\tfrac\ell{\alpha}\right)\ 
q^{n/2} r^{\ell/2} s^{m/2}\ .
\end{equation}

\section{Explicit formulae for $\Delta_{k/2}(\mathbf{Z})$}

We note that $\Delta_{k/2}(\mathbf{Z})$ is symmetric under the exchange
$z_1\leftrightarrow z_3$ and is anti-symmetric under $z_2\rightarrow -z_2$
for all values of $k$.
\begin{eqnarray*}
 \Delta_5&=&\left( - \frac{1}{{\sqrt{r}}} 
         + {\sqrt{r}} \right) \,\sqrt{q}{\sqrt{s}} + 
 \left( \frac{9}{r^{\frac{5}{2}}} - 
     \frac{93}{r^{\frac{3}{2}}} + 
     \frac{90}{{\sqrt{r}}} - 90\,{\sqrt{r}} + 
     93\,r^{\frac{3}{2}} - 9\,r^{\frac{5}{2}} \right) \,
   q^{\frac{3}{2}} s^{\frac{3}{2}} \\
   &+& \left( r^{- \frac{3}{2} } + 
     \frac{9}{{\sqrt{r}}} - 9\,{\sqrt{r}} - 
     r^{\frac{3}{2}} \right) \,
   \left( q^{\frac{3}{2}}\,{\sqrt{s}} + 
     {\sqrt{q}}\,s^{\frac{3}{2}} \right)  \\ 
     &+ &
  \left( \frac{-9}{r^{\frac{3}{2}}} - 
     \frac{27}{{\sqrt{r}}} + 27\,{\sqrt{r}} + 
     9\,r^{\frac{3}{2}} \right) \,
   \left( q^{\frac{5}{2}}\,{\sqrt{s}} + 
     {\sqrt{q}}\,s^{\frac{5}{2}} \right)  \\
     &+& 
  \left( -r^{- \frac{5}{2}  } + 
     \frac{27}{r^{\frac{3}{2}}} + 
     \frac{12}{{\sqrt{r}}} - 12\,{\sqrt{r}} - 
     27\,r^{\frac{3}{2}} + r^{\frac{5}{2}} \right) \,
   \left( q^{\frac{7}{2}}\,{\sqrt{s}} + 
     {\sqrt{q}}\,s^{\frac{7}{2}} \right) \\ 
     &+& 
  \left( \frac{9}{r^{\frac{5}{2}}} - 
     \frac{12}{r^{\frac{3}{2}}} + 
     \frac{90}{{\sqrt{r}}} - 90\,{\sqrt{r}} + 
     12\,r^{\frac{3}{2}} - 9\,r^{\frac{5}{2}} \right) \,
   \left( q^{\frac{9}{2}}\,{\sqrt{s}} + 
     {\sqrt{q}}\,s^{\frac{9}{2}} \right) \\ 
     &+ &
  \left( \frac{-27}{r^{\frac{5}{2}}} - 
     \frac{90}{r^{\frac{3}{2}}} - 
     \frac{135}{{\sqrt{r}}} + 135\,{\sqrt{r}} + 
     90\,r^{\frac{3}{2}} + 27\,r^{\frac{5}{2}} \right) \,
   \left( q^{\frac{11}{2}}\,{\sqrt{s}} + 
     {\sqrt{q}}\,s^{\frac{11}{2}} \right)  \\
     &+&
  \left( r^{- \frac{7}{2}  } + 
     \frac{12}{r^{\frac{5}{2}}} + 
     \frac{135}{r^{\frac{3}{2}}} - 
     \frac{54}{{\sqrt{r}}} + 54\,{\sqrt{r}} - 
     135\,r^{\frac{3}{2}} - 12\,r^{\frac{5}{2}} - 
     r^{\frac{7}{2}} \right) \,
   \left( q^{\frac{13}{2}}\,{\sqrt{s}} + 
     {\sqrt{q}}\,s^{\frac{13}{2}} \right)  
%      \\+ 
%   \left( \frac{-9}{r^{\frac{7}{2}}} + 
%      \frac{90}{r^{\frac{5}{2}}} + 
%      \frac{54}{r^{\frac{3}{2}}} + 
%      \frac{99}{{\sqrt{r}}} - 99\,{\sqrt{r}} - 
%      54\,r^{\tfrac{3}{2}} - 90\,r^{\tfrac{5}{2}} + 
%      9\,r^{\tfrac{7}{2}} \right) \,
%    \left( q^{\tfrac{15}{2}}\,s^{\tfrac12} + 
%      q^{\tfrac{5}{2}}\,s^{\tfrac{3}{2}} + 
%      q^{\tfrac{3}{2}}\,s^{\tfrac{5}{2}} + 
%      q^{\tfrac12}\,s^{\tfrac{15}{2}} \right)
\end{eqnarray*}
\begin{eqnarray*}
 \Delta_3&=&\left( - \frac{1}{{\sqrt{r}}} 
         + {\sqrt{r}} \right) \,\sqrt{q}{\sqrt{s}} + 
  \left( r^{- \frac{5}{2}  } - 
     \frac{5}{r^{\frac{3}{2}}} - \frac{6}{{\sqrt{r}}} + 
     6\,{\sqrt{r}} + 5\,r^{\frac{3}{2}} - r^{\frac{5}{2}}
     \right) \,q^{\frac{3}{2}}s^{\frac{3}{2}} \\
     &+& 
  \left( r^{- \frac{3}{2}  } + 
     \frac{1}{{\sqrt{r}}} - {\sqrt{r}} - r^{\frac{3}{2}}
     \right) \,\left( q^{\frac{3}{2}}\,{\sqrt{s}} + 
     {\sqrt{q}}\,s^{\frac{3}{2}} \right)  \\
     &+& 
  \left( -r^{- \frac{3}{2}  } + 
     \frac{5}{{\sqrt{r}}} - 5\,{\sqrt{r}} + 
     r^{\frac{3}{2}} \right) \,
   \left( q^{\frac{5}{2}}\,{\sqrt{s}} + 
     {\sqrt{q}}\,s^{\frac{5}{2}} \right) \\ 
     &+& 
  \left( -r^{-\frac{5}{2}  } - 
     \frac{5}{r^{\frac{3}{2}}} - \frac{4}{{\sqrt{r}}} + 
     4\,{\sqrt{r}} + 5\,r^{\frac{3}{2}} + r^{\frac{5}{2}}
     \right) \,\left( q^{\frac{7}{2}}\,{\sqrt{s}} + 
     {\sqrt{q}}\,s^{\frac{7}{2}} \right)  \\
     &+& 
  \left( r^{- \frac{5}{2}  } + 
     \frac{4}{r^{\frac{3}{2}}} - \frac{6}{{\sqrt{r}}} + 
     6\,{\sqrt{r}} - 4\,r^{\frac{3}{2}} - r^{\frac{5}{2}}
     \right) \,\left( q^{\frac{9}{2}}\,{\sqrt{s}} + 
     {\sqrt{q}}\,s^{\frac{9}{2}} \right)\\  
     &+& 
  \left( \frac{5}{r^{\frac{5}{2}}} + 
     \frac{6}{r^{\frac{3}{2}}} + \frac{1}{{\sqrt{r}}} - 
     {\sqrt{r}} - 6\,r^{\frac{3}{2}} - 5\,r^{\frac{5}{2}}
     \right) \,\left( q^{\frac{11}{2}}\,{\sqrt{s}} + 
     {\sqrt{q}}\,s^{\frac{11}{2}} \right) \\ 
     &+& 
  \left( r^{-\frac{7}{2}  } - 
     \frac{4}{r^{\frac{5}{2}}} - 
     r^{-\left( \frac{3}{2} \right) } - 
     \frac{6}{{\sqrt{r}}} + 6\,{\sqrt{r}} + 
     r^{\frac{3}{2}} + 4\,r^{\frac{5}{2}} - 
     r^{\frac{7}{2}} \right) \,
   \left( q^{\frac{13}{2}}\,{\sqrt{s}} + 
     {\sqrt{q}}\,s^{\frac{13}{2}} \right)  
%      \\+ 
%   \left( -r^{-\left( \frac{7}{2} \right) } - 
%      \frac{6}{r^{\frac{5}{2}}} + 
%      \frac{6}{r^{\frac{3}{2}}} + \frac{11}{{\sqrt{r}}} - 
%      11\,{\sqrt{r}} - 6\,r^{\frac{3}{2}} + 
%      6\,r^{\frac{5}{2}} + r^{\frac{7}{2}} \right) \,
%    \left( q^{\frac{15}{2}}\,{\sqrt{s}} + 
%      q^{\frac{5}{2}}\,s^{\frac{3}{2}} + 
%      q^{\frac{3}{2}}\,s^{\frac{5}{2}} + 
%      {\sqrt{q}}\,s^{\frac{15}{2}} \right)
\end{eqnarray*}

\begin{eqnarray*}
\Delta_2&=&
   \left(\sqrt{r}-\frac{1}{\sqrt{r}}\right)
 \sqrt{q}  \sqrt{s}+ \left(3 r^{3/2}-\frac{3}{r^{3/2}}\right)
   q^{3/2} s^{3/2}+\left(\frac{1}{r^{3/2}}-r^{3/2}\right)
   \left(\sqrt{s} q^{3/2}+s^{3/2}
   \sqrt{q}\right)\\
   &+&\left(r^{5/2}-3
   \sqrt{r}+\frac{3}{\sqrt{r}}-\frac{1}{r^{5/2}}\right)
   \left(\sqrt{s} q^{7/2}+s^{7/2} \sqrt{q}\right)+\left(3
   r^{3/2}-\frac{3}{r^{3/2}}\right) \left(\sqrt{s}
   q^{9/2}+s^{9/2} \sqrt{q}\right)\\
   &+&\left(-r^{7/2}-3
   r^{5/2}+\frac{3}{r^{5/2}}+\frac{1}{r^{7/2}}\right)
   \left(\sqrt{s} q^{13/2}+s^{13/2}
   \sqrt{q}\right)\\
   &+&\left(3 r^{7/2}+5
   \sqrt{r}-\frac{5}{\sqrt{r}}-\frac{3}{r^{7/2}}\right)
   \left(\sqrt{s} q^{19/2}+s^{19/2} \sqrt{q}\right)
\end{eqnarray*}
\begin{eqnarray*}
\Delta_1&=&\left(\sqrt{r}-\frac{1}{\sqrt{r}}\right)
  \sqrt{q} \sqrt{s}
+\left(-r^{3/2}+\sqrt{r}-\frac{1}{\sqrt{r}}+\frac{1}{r^{3/2}}\right)
\left(\sqrt{s} q^{3/2}+s^{3/2} \sqrt{q}\right)\\
&+& \left(r^{5/2}-2 r^{3/2}+5
   \sqrt{r}-\frac{5}{\sqrt{r}}+\frac{2}{r^{3/2}}-\frac{1}{
   r^{5/2}}\right) q^{3/2}s^{3/2} \\
 &+&  \left(-r^{3/2}+2
   \sqrt{r}-\frac{2}{\sqrt{r}}+\frac{1}{r^{3/2}}\right)
   \left(\sqrt{s} q^{5/2}+s^{5/2}
   \sqrt{q}\right)\\
   &+&
   \left(r^{5/2}-2 r^{3/2}+3
   \sqrt{r}-\frac{3}{\sqrt{r}}+\frac{2}{r^{3/2}}-\frac{1}{
   r^{5/2}}\right) \left(\sqrt{s} q^{7/2}+s^{7/2}
   \sqrt{q}\right)\\
   &+&\left(r^{5/2}-3 r^{3/2}+5
   \sqrt{r}-\frac{5}{\sqrt{r}}+\frac{3}{r^{3/2}}-\frac{1}{
   r^{5/2}}\right) \left(\sqrt{s} q^{9/2}+s^{9/2}
   \sqrt{q}\right)\\
   &+&\left(2 r^{5/2}-5 r^{3/2}+5
   \sqrt{r}-\frac{5}{\sqrt{r}}+\frac{5}{r^{3/2}}-\frac{2}{
   r^{5/2}}\right) \left(\sqrt{s} q^{11/2}+s^{11/2}
   \sqrt{q}\right)\\
   &+&\left(-r^{7/2}+3 r^{5/2}-5 r^{3/2}+9
   \sqrt{r}-\frac{9}{\sqrt{r}}+\frac{5}{r^{3/2}}-\frac{3}{
   r^{5/2}}+\frac{1}{r^{7/2}}\right) \left(\sqrt{s}
   q^{13/2}+s^{13/2} \sqrt{q}\right)
%\\
%&-&\left(r^{7/2}-5
%r^{5/2}+9 r^{3/2}-11
%\sqrt{r}+\frac{11}{\sqrt{r}}-\frac{9}{r^{3/2}}+\frac{5}
%{r^{5/2}}-\frac{1}{r^{7/2}}\right) \left(\sqrt{s}
%q^{15/2}+s^{3/2} q^{5/2}+s^{5/2} q^{3/2}+s^{15/2}
%\sqrt{q}\right)
\end{eqnarray*}
\subsection{Analyzing the modular forms}
We list some of the observations that can be made from the above expansions.
\begin{enumerate}
\item Using the expressions for the real simple roots, 
$(\delta_1,\delta_2,\delta_3)$ and their inner product with 
$\mathbf{Z}$, one sees that
$$
e^{-\pi i (\delta_1,\mathbf{Z})}=q r\ ,\  e^{-\pi i (\delta_2,\mathbf{Z})}=r^{-1} \  
\textrm{ and }e^{-\pi i (\delta_3,\mathbf{Z})}=s r\ .
$$
(Recall that $q=\exp(2\pi i z_1)$, $r=\exp(2\pi i z_2)$ and 
$s=\exp(2\pi i z_3)$.) Thus, one has $\exp(-\pi i 
(\rho,\mathbf{Z}))=q^{1/2} r^{1/2} s^{1/2}$. Further, one has the 
identification relating the root $\alpha[n,\ell,m]$ to $q^n 
r^\ell s^m$:
$$
q^n r^\ell s^m = e^{-\pi i (\alpha[n,\ell,m], \mathbf{Z})}\ ,
$$
where the root $\alpha[n,\ell,m]=n \delta_1 + (-l+m+n)\delta_2 + 
m \delta_3$ has norm $(2\ell^2-8nm)$. The real simple roots are 
$(\alpha[1,1,0],\alpha[0,-1,0],\alpha[0,1,1])$ and the Weyl 
vector is $\rho=\alpha[\tfrac12,\tfrac12,\tfrac12]$ in this 
notation.
\item In the expansion for $\Delta_5(\mathbf{Z})$, all terms (in 
the expansion given above) that arise with coefficient $\pm1$ 
come from the sum side of the Lie algebra $\Fg(A_{1,II})$. In 
other words, they arise by the action of all elements of the Weyl 
group generated by the three real simple roots. For instance, the 
terms arising from Weyl reflections associated with the simple 
real roots of $\mathcal{G}_1$ are
$$
(q^{3/2}r^{3/2}s^{1/2}, q^{1/2}r^{-1/2}s^{1/2}, q^{1/2}r^{3/2}s^{3/2})=q^{1/2}r^{1/2}s^{1/2}(qr,r^{-1},sr)\ .
$$ 
Note that we need to pull out an overall factor of 
$q^{1/2}r^{1/2}s^{1/2}$ in the sum side of the denominator 
formula to extract the roots.
\item The terms that appear in $\Delta_5(\mathbf{Z})$ with 
coefficient $\pm1$ continue do so in the other modular forms 
$\Delta_{k/2}(\mathbf{Z})$. This is consistent with the 
observation that the real roots are unaffected by the 
orbifolding. We have also verified that the terms involving 
imaginary roots related by Weyl reflections appear with the same 
multiplicity.
\item All the GKM superalgebras have an outer $S_3$ symmetry 
which permutes the three real simple roots. It is easy to see 
only the $\delta_1\leftrightarrow \delta_3$ (or equivalently the 
$q\leftrightarrow s$) symmetry in the $\Delta_{k/2}(\mathbf{Z})$. 
A formal proof can be given by following Gritsenko and Nikulin's 
argument for $\mathcal{G}_1$\cite[see Prop. 2.1]{Nikulin:1995}. 
Their proof makes use of the non-trivial character $v^\Gamma$ 
appearing in the modular transform $\Delta_5(\mathbf{Z})$. All 
the modular forms, $\Delta_{k/2}(\mathbf{Z})$, share the same 
character suitably restricted to the relevant subgroup of 
$Sp(2,\BZ)$, the same proof goes through.
\item A practical check of the outer $S_3$ needs us to verify the 
$\delta_1\leftrightarrow \delta_2$ invariance of 
$\Delta_k(\mathbf{Z})$. One can show that under this exchange
$$
\alpha[n,\ell,m] \leftrightarrow \alpha[-\ell+m+n,-\ell+2m,m]\ .
$$ 
For instance, the light-like root $\alpha[0,0,1]$ is mapped to 
another light-like root $\alpha[1,2,1]$. This relates the term 
$q^{1/2}r^{1/2}s^{3/2}$ to $q^{3/2}r^{5/2}s^{3/2}$ -- both have 
multiplicity $-9$ in $\Delta_5(\mathbf{Z})$. These two terms 
appear with identical multiplicity $-1$ in 
$\Delta_3(\mathbf{Z})$, $0$ in $\Delta_2(\mathbf{Z})$ and $1$ in 
$\Delta_1(\mathbf{Z})$. A similar analysis has been carried out 
for other terms as well.
\item Another proof of the outer $S_3$ symmetry can be obtained 
by considering the product form of the modular form in Eq. 
\eqref{productformdeltak}. The multiplicities of positive roots 
is determined by two functions $c_1(nm,\ell)$ and $c_2(nm,\ell)$. 
These functions depend only on the combination $(4nm-\ell^2)$ as 
they arise from the Fourier expansions of weak Jacobi forms. The 
action of the outer $S_3$ maps a positive root 
$\alpha[n,\ell,m]$, with norm $-2(4nm-\ell^2)$, to another root 
with the same norm.  Both these roots necessarily share the same 
multiplicity.
\end{enumerate}
 

\begin{thebibliography}{99}
%\cite{Strominger:1996sh}
\bibitem{Strominger:1996sh}
A.~Strominger and C.~Vafa,
``Microscopic Origin of the Bekenstein-Hawking Entropy,''
Phys.\ Lett.\  B {\bf 379} (1996) 99
{\tt arXiv:hep-th/9601029}.
%%CITATION = PHLTA,B379,99;%%
%\cite{Duff:1995sm}
\bibitem{Duff:1995sm}
M.~J.~Duff, J.~T.~Liu and J.~Rahmfeld,
``Four-Dimensional String-String-String Triality,''
Nucl.\ Phys.\  B {\bf 459} (1996) 125
[arXiv:hep-th/9508094].
%%CITATION = NUPHA,B459,125;%%
%\cite{Dijkgraaf:1996it}
\bibitem{Dijkgraaf:1996it}
R.~Dijkgraaf, E.~P.~Verlinde and H.~L.~Verlinde,
``Counting dyons in N = 4 string theory,''
Nucl.\ Phys.\  B {\bf 484} (1997) 543
{\tt arXiv:hep-th/9607026}.
%%CITATION = NUPHA,B484,543;%%
\bibitem{Nikulin:1995}
V. A. Gritsenko, V. V. Nikulin,
``Siegel Automorphic Form Corrections of Some Lorentzian 
Kac-Moody Lie Algebras,''
Amer. J. Math.  \textbf{119} (1997) 181-224
{\tt arXiv:alg-geom/9504006} 
%\cite{Harvey:1995fq}
\bibitem{Harvey:1995fq}
J.~A.~Harvey and G.~W.~Moore,
``Algebras, BPS States, and Strings,''
Nucl.\ Phys.\  B {\bf 463} (1996) 315
{\tt arXiv:hep-th/9510182}.
%%CITATION = NUPHA,B463,315;%%
%\cite{Harvey:1996gc}
\bibitem{Harvey:1996gc}
J.~A.~Harvey and G.~W.~Moore,
``On the algebras of BPS states,''
Commun.\ Math.\ Phys.\  {\bf 197} (1998) 489
{\tt arXiv:hep-th/9609017}.
%%CITATION = CMPHA,197,489;%%
%\cite{Cheng:2008fc}
\bibitem{Cheng:2008fc}
M.~C.~N.~Cheng and E.~P.~Verlinde,
``Wall Crossing, Discrete Attractor Flow, and Borcherds Algebra,''
SIGMA \textbf{4} (2008) 068 
{\tt arXiv:0806.2337 [hep-th]}.
%%CITATION = ARXIV:0806.2337;%%
%\cite{Chaudhuri:1995fk}
\bibitem{Chaudhuri:1995fk}
S.~Chaudhuri, G.~Hockney and J.~D.~Lykken,
``Maximally Supersymmetric String Theories in $D < 10$,''
Phys.\ Rev.\ Lett.\  {\bf 75} (1995) 2264
{\tt arXiv:hep-th/9505054}.
%%CITATION = PRLTA,75,2264;%%
%\cite{Jatkar:2005bh}
\bibitem{Jatkar:2005bh}
 D.~P.~Jatkar and A.~Sen,
``Dyon spectrum in CHL models,''
JHEP {\bf 0604}, 018 (2006)
{\tt arXiv:hep-th/0510147}.
%%CITATION = JHEPA,0604,018;%%
%\cite{Sen:2005pu}
\bibitem{Sen:2005pu}
A.~Sen,
``Black Holes, Elementary Strings and Holomorphic Anomaly,''
JHEP {\bf 0507} (2005) 063
{\tt arXiv:hep-th/0502126}.
%%CITATION = JHEPA,0507,063;%%
\bibitem{Borcherds:1990}
R.~Borcherds,
 ``The Monster Lie algebra,''
Advances \ In \ Mathematics {\bf 83}, 30-47 (1990)
%%%CITATION = NONE;%%
\bibitem{Niemann} P.~Niemann,
``Some generalized Kac-Moody algebras with known root multiplicities,''
Mem. Amer. Math. Soc. 157 (2002), no. 746 
{\tt arXiv:math.QA/0001029}
%%%CITATION = NONE;%%
%\cite{Schwarz:1993vs}
\bibitem{Schwarz:1993vs}
J.~H.~Schwarz and A.~Sen,
``Duality symmetric actions,''
Nucl.\ Phys.\  B {\bf 411} (1994) 35
{\tt arXiv:hep-th/9304154}.
%%CITATION = NUPHA,B411,35;%%
%\cite{Banerjee:2007sr}
\bibitem{Banerjee:2007sr}
S.~Banerjee and A.~Sen,
``Duality Orbits, Dyon Spectrum and Gauge Theory Limit of Heterotic String
Theory on $T^6$,''
JHEP {\bf 0803} (2008) 022
{\tt arXiv:0712.0043 [hep-th]}.
%%CITATION = JHEPA,0803,022;%%
%\cite{Cvetic:1995uj}
\bibitem{Cvetic:1995uj}
M.~Cvetic and D.~Youm,
``Dyonic BPS saturated black holes of heterotic string on a six torus,''
Phys.\ Rev.\  D {\bf 53} (1996) 584
{\tt arXiv:hep-th/9507090}.
%%CITATION = PHRVA,D53,584;%%
%\cite{Lerche:1999ju}
\bibitem{Lerche:1999ju}
W.~Lerche and S.~Stieberger,
``1/4 BPS states and non-perturbative couplings in N = 4 string theories,''
Adv.\ Theor.\ Math.\ Phys.\  {\bf 3} (1999) 1539
{\tt arXiv:hep-th/9907133}.
%%CITATION = 00203,3,1539;%%
%\cite{David:2006ji}
\bibitem{David:2006ji}
J.~R.~David, D.~P.~Jatkar and A.~Sen,
``Product representation of dyon partition function in CHL models,''
JHEP {\bf 0606}, 064 (2006)
{\tt arXiv:hep-th/0602254}.
%%CITATION = JHEPA,0606,064;%%
%\cite{Dabholkar:2005by}
\bibitem{Dabholkar:2005by}
A.~Dabholkar, F.~Denef, G.~W.~Moore and B.~Pioline,
``Exact and asymptotic degeneracies of small black holes,''
JHEP {\bf 0508} (2005) 021
{\tt arXiv:hep-th/0502157}.
%%CITATION = JHEPA,0508,021;%%
%\cite{Dabholkar:2005dt}
\bibitem{Dabholkar:2005dt}
A.~Dabholkar, F.~Denef, G.~W.~Moore and B.~Pioline,
``Precision counting of small black holes,''
JHEP {\bf 0510} (2005) 096
{\tt arXiv:hep-th/0507014}.
%%CITATION = JHEPA,0510,096;%%
%\cite{Sen:2005iz}
%\cite{Dabholkar:2004yr}
\bibitem{Dabholkar:2004yr}
A.~Dabholkar,
``Exact counting of black hole microstates,''
Phys.\ Rev.\ Lett.\  {\bf 94} (2005) 241301
{\tt arXiv:hep-th/0409148}.
%%CITATION = PRLTA,94,241301;%%
\bibitem{Sen:2005iz}
A.~Sen,
``Entropy function for heterotic black holes,''
JHEP {\bf 0603} (2006) 008
{\tt arXiv:hep-th/0508042}.
%%CITATION = JHEPA,0603,008;%%
%\cite{Dabholkar:2006xa}
\bibitem{Dabholkar:2006xa}
A.~Dabholkar and S.~Nampuri,
``Spectrum of Dyons and Black Holes in CHL orbifolds using Borcherds Lift,''
JHEP {\bf 0711}, 077 (2007)
{\tt arXiv:hep-th/0603066}.
%%CITATION = JHEPA,0711,077;%%
%\cite{David:2006yn}
\bibitem{David:2006yn}
J.~R.~David and A.~Sen,
``CHL dyons and statistical entropy function from D1-D5 system,''
JHEP {\bf 0611} (2006) 072
{\tt arXiv:hep-th/0605210}.
%%CITATION = JHEPA,0611,072;%%
%\cite{Shih:2005uc}
\bibitem{Shih:2005uc}
D.~Shih, A.~Strominger and X.~Yin,
``Recounting dyons in N = 4 string theory,''
JHEP {\bf 0610}, 087 (2006)
{\tt arXiv:hep-th/0505094}.
%%CITATION = JHEPA,0610,087;%%
%\cite{Dijkgraaf:1996xw}
\bibitem{Dijkgraaf:1996xw}
R.~Dijkgraaf, G.~W.~Moore, E.~P.~Verlinde and H.~L.~Verlinde,
``Elliptic genera of symmetric products and second quantized strings,''
Commun.\ Math.\ Phys.\  {\bf 185} (1997) 197
{\tt arXiv:hep-th/9608096}.
%%CITATION = CMPHA,185,197;%%
\bibitem{Kac:1990}
V. G. Kac,
``Infinite dimensional Lie algebras, 3rd ed.,''
Cambridge \ University \ Press, (1990).
%%%CITATION = NONE;%%
\bibitem{Feingold:1983}
A. J. Feingold and I. B. Frenkel, 
``A hyperbolic Kac-Moody algebra and the theory of Siegel modular 
forms of genus 2', 
Math. Ann. \textbf{263} (1983), 87-144.
%%%CITATION = NONE;%%
%\cite{Feingold:2003es}
\bibitem{Feingold:2003es}
A.~J.~Feingold and H.~Nicolai,
``Subalgebras of Hyperbolic Kac-Moody Algebras,''
{\tt arXiv:math/0303179}.
%%CITATION = MATH/0303179;%%
\bibitem{Lepowsky:1978}
J. Lepowsky, S. Milne.
``Lie Algebraic approaches to classical partition identities,''
Advances \ In \ Mathematics {\bf 29}, (1978) 15-29.
\bibitem{MacDonald:1972} I.~G.~MacDonald, ``Affine root systems and Dedekind's
$\eta$-function,'' Invent. Math \textbf{15} (1972) 91-143.
%%%CITATION = NONE;%%
\bibitem{Lepowsky:1976}
J. Lepowsky, H. Garland,
``Lie Algebra homology and the Macdonald-Kac formulas,''
Invent. \ Math. {\bf 34}, (1976) 37-76.
%%%CITATION = NONE;%%
\bibitem{Borcherds:1988pq}
R.~Borcherds,
``Generalized Kac-Moody algebras,'' 
Journal \ of \ Algebra {\bf 115}, 501-512 (1988)
%%%CITATION = NONE;%%
\bibitem{Borcherds:1992}
R.~Borcherds,
``Monstrous Moonshine and Monstrous Lie Superalgebras,''
Invent. \ Math. {\bf 109}, (1992) 405-444.
%%%CITATION = NONE;%%
\bibitem{Jurisich:1996}
E. Jurisich,
``An exposition of Generalized Kac-Moody algebras.''
Contemporary \ Mathematics vol. {\bf 194}, (1996) 121-159.
%%%CITATION = NONE;%%
\bibitem{Aoki:2005}
H.~Aoki and T.~Ibukiyama, ``Simple Graded Rings of Siegel Modular 
Forms, Differential Operators and Borcherds Products,'' 
Int. J. Math. {\bf 16} 3 (2005), 249-279.
%%%CITATION = NONE;%%
%\cite{Dabholkar:2006bj}
\bibitem{Dabholkar:2006bj}
A.~Dabholkar and D.~Gaiotto,
 ``Spectrum of CHL dyons from genus-two partition function,''
JHEP {\bf 0712} (2007) 087
{\tt arXiv:hep-th/0612011}.
%%CITATION = JHEPA,0712,087;%%
%\cite{Sen:2007vb}
\bibitem{Sen:2007vb}
A.~Sen,
``Walls of Marginal Stability and Dyon Spectrum in N=4 Supersymmetric
String Theories,''
JHEP {\bf 0705} (2007) 039
{\tt arXiv:hep-th/0702141}.
%%CITATION = JHEPA,0705,039;%%
%\cite{Sen:2007nz}
\bibitem{Sen:2007nz}
A.~Sen,
``Rare Decay Modes of Quarter BPS Dyons,''
JHEP {\bf 0710} (2007) 059
{\tt arXiv:0707.1563 [hep-th]}.
%%CITATION = JHEPA,0710,059;%%
%\cite{Mukherjee:2007nc}
\bibitem{Mukherjee:2007nc}
A.~Mukherjee, S.~Mukhi and R.~Nigam,
``Dyon Death Eaters,''
JHEP {\bf 0710} (2007) 037
{\tt arXiv:0707.3035 [hep-th]}.
%%CITATION = JHEPA,0710,037;%%
%\cite{Mukherjee:2007af}
\bibitem{Mukherjee:2007af}
A.~Mukherjee, S.~Mukhi and R.~Nigam,
``Kinematical Analogy for Marginal Dyon Decay,''
{\tt arXiv:0710.4533 [hep-th]}.
%%CITATION = ARXIV:0710.4533;%%
%\cite{Cheng:2008kt}
\bibitem{Cheng:2008kt}
M.~C.~N.~Cheng and A.~Dabholkar,
``Borcherds-Kac-Moody Symmetry of $\mathcal{N}=4$ Dyons,''
{\tt arXiv:0809.4258 [hep-th]}.
%%CITATION = ARXIV:0809.4258;%%
\bibitem{Nikulin:1996}
V. A. Gritsenko, V. V. Nikulin,
``The Igusa Modular Forms and ``the simples" Lorentzian Kac-Moody algebra,''
(Russian) Mat. Sb.  \textbf{187}  (1996),  no. 11, 27--66;  translation in  Sb. Math.  \textbf{187}  (1996),  no. 11, 1601--1641
{\tt arXiv:alg-geom/9603010}.
\bibitem{Garbagnati:2006}
A.~Garbagnati and A.~Sarti,
``Symplectic automorphisms of prime order on K3 surfaces,"
J. of Algebra \textbf{318} (2007) 323-350
 {\tt arXiv:math/0603742}.
%%%CITATION = NONE;%%
\bibitem{Garbagnati:2008}
A.~Garbagnati and A.~Sarti,
`` Elliptic fibrations and symplectic automorphisms on K3 surfaces," 
{\tt arXiv:0801.3992v1 [math.AG]}.
%%%CITATION = NONE;%%
\bibitem{Gritsenko:1999} V.~Gritsenko, 
``Elliptic genus of Calabi-Yau manifolds and Jacobi and Siegel modular forms,''
 Algebra i Analiz 11 (1999), no. 5, 100--125; translation in
St. Petersburg Math. J. 11 (2000), no. 5, 781--804 
{\tt arXiv:math/9906190v1}.
%%%CITATION = NONE;%%
%\cite{Gopakumar:1998ii}
\bibitem{Gopakumar:1998ii}
R.~Gopakumar and C.~Vafa,
 ``M-theory and topological strings. I,''
{\tt arXiv:hep-th/9809187}.
%%CITATION = HEP-TH/9809187;%%
%\cite{Gopakumar:1998jq}
\bibitem{Gopakumar:1998jq}
R.~Gopakumar and C.~Vafa,
``M-theory and topological strings. II,''
{\tt arXiv:hep-th/9812127}.
%%CITATION = HEP-TH/9812127;%%
\bibitem{Young:2008}
B.~Young (with an appendix by J.~Bryan),
``Generating functions for colored 3D Young diagrams and the
Donaldson-Thomas invariants of orbifolds,''
{\tt arXiv:0802.2948[math.CO]}.
%%%CITATION = NONE;%%
%\bibitem{Moody:1968}
%R. Moody,
%``A new class of Lie algebras,''
%Journal \ of \ Algebra {\bf 10}, (1968) 211-230.
\bibitem{Raghavan}
S.~Raghavan,
``Cusp forms of degree $2$ and weight $3$,''
Math. Ann. {\bf 224}, no. 2 (1976), 149--156. 
%%%CITATION = NONE;%%
\bibitem{Eichler} M.~Eichler and D.~Zagier, ``The theory of Jacobi Forms," Birkh\"auser (1985).
%%%CITATION = NONE;%%
\bibitem{Maass} H.~Maa\ss,Inv. Math. \textbf{52} (1979) 95-104, Ibid. \textbf{53} (1979) 249-253, Ibid. \textbf{53} (179) 255-265.
%%%CITATION = NONE;%%
\bibitem{Dirichletchar} We found it useful to refer to the online
encyclopedias:\qquad\qquad\mbox{}
\texttt{http://planetmath.org/encyclopedia/DirichletCharacter.html} as well as
\texttt{http://en.wikipedia.org/wiki/Dirichlet\_character}.
%%%CITATION = NONE;%%
\bibitem{Reiner:1955}
 I.~Reiner, 
 ``Real linear characters of the symplectic modular group," 
 Proc. Amer. Math. Soc. \textbf{6} (1955) pp. 987-990.\\
%%%CITATION = NONE;%%
 H.~Maa\ss, ``Die Multiplikatorsysteme zur Siegelschen Modulgruppe''  Nachr.
Akad. Wiss.G\"ottingen Math.-Phys. Kl. II  (1964) 
125--135.
%%%CITATION = NONE;%%
\bibitem{Manickam:1993} M. Manickam, B. Ramakrishnan, ``On Saito-Kurokawa Descent for
congruence subgroups," Manuscripta Math. \textbf{81} (1993) 161-182.
\bibitem{Manickam:2002} M. Manickam, B. Ramakrishnan, ``On Saito-Kurokawa correspondence of degree two for arbitrary level," J. Ramanujan Math. Soc. \textbf{17-3} (2002) 149-160.
\bibitem{Aoki:2008} H.~Aoki, ``A note on additive lifts of Jacobi forms with levels" (under preparation).
\bibitem{Maass:1980} H.~Maa\ss, ``\"Uber ein Analogon zur Vermutung yon Saito-Kurokawa," Inventiones. Math. \textbf{60} 1980 85-104.
%%%CITATION = NONE;%%
\end{thebibliography}
\end{document}